\documentclass{llncs}

\usepackage{epsfig}
\usepackage{llncsdoc}

\def\C{{\cal C}} 
\def\S{{\cal S}} 
\def\D{{\cal D}} 
\def\M{{\cal M}} 
\def\T{{\cal T}} 
\def\fmind{{\cal FM-index}}
\def\wfmind{{\cal WFM-index}}
\def\suf{SU\!F}

\newcommand{\hk}[1]{H_{#1}}    
\newcommand{\alg}[1]{\bigbreak\noindent{\bf #1}}
\newcommand{\algskip}{\itemsep=-3pt\baselineskip=12pt}
\newcommand{\bzip}{{\tt bzip2}}
\newcommand{\gzip}{{\tt gzip}}
\newcommand{\countx}{{\sf get\_rows}}
\newcommand{\countxx}{{\sf count}}
\newcommand{\locate}{{\sf locate}}
\newcommand{\num}{\mbox{\sf Occ}}
\newcommand{\ch}{{c}}

\title{\bf String algorithms and data structures\thanks{Address:
Dipartimento di Informatica, Corso Italia 40, 56125 Pisa, Italy,
\email{ferragina@di.unipi.it}, \texttt{http://www.di.unipi.it/$\sim$ferragin}.
Partially supported by Italian MIUR projects: ``Technologies and
services for enhanced content delivery'' and ``A high-performance
distributed platform''.}}

\author{Paolo Ferragina} 

\tocauthor{Paolo Ferragina (University of Pisa, Italy)}
\institute{Dipartimento di Informatica, Universit\`a di Pisa,
Italy}

\date{}

\authorrunning{Paolo Ferragina}
\titlerunning{String algorithms and data structures}

\begin{document}

\maketitle

%%%%%%%%%% ABSTRACT %%%%%%%%%%%%%%%%%%%%%%%%%%
\begin{abstract}

The string-matching field has grown at a such complicated stage that
various issues come into play when studying it: data structure and
algorithmic design, database principles, compression techniques,
architectural features, cache and prefetching policies. The expertise
nowadays required to design good string data structures and algorithms
is therefore transversal to many computer science fields and much more
study on the orchestration of known, or novel, techniques is needed to
make progress in this fascinating topic.  This survey is aimed at
illustrating the key ideas which should constitute, in our opinion,
the current background of every {\em index designer}. We also discuss
the positive features and drawbacks of known indexing schemes and
algorithms, and devote much attention to detail research issues and
open problems both on the theoretical and the experimental side.

\end{abstract}

%%%%%%%%%% INTRODUCTION %%%%%%%%%%%%%%%%%%%%%%%%%%

\section{Introduction} \label{intro}

String data is ubiquitous, common-place applications are digital
libraries and product catalogs (for books, music, software, etc.),
electronic white and yellow page directories, specialized information
sources (e.g. patent or genomic databases), customer relationship
management of data, etc.. The amount of textual information managed by
these applications is increasing at a staggering rate. The best two
illustrative examples of this growth are the World-Wide Web, which is
estimated to provide access to at least three terabytes of textual
data, and the genomic databases, which are estimated to store more
than fifteen billion of base pairs. Even in private hands are common
now collection sizes which were unimaginable a few years ago.

This scenario is destined to become more pervasive due to the
migration of current databases toward XML storage~\cite{xml-w3c}. XML
is emerging as the {\em de facto} standard for the publication and
interchange of heterogeneous, incomplete and irregular data over the
Internet and amongst applications. It provides ground rules to mark up
data so it is self-describing and easily readable by humans and
computers. Large portions of XML data are textual and include
descriptive fields and tags.  Evaluating an XML query involves
navigating paths through a tree (or, in general, a graph)
structure. In order to speed up query processing, current approaches
consist of encoding document paths into strings of arbitrary length
(e.g. {\tt book/author/firstname/}) and replacing tree navigational
operations with string prefix queries (see
e.g.~\cite{fabric,lore1,aboulnaga}).

In all these situations brute-force scanning of such large collections
is not a viable approach to perform string searches. Some kind of {\em
index} has to be necessarily built over these massive textual data to
effectively process string queries (of arbitrarily lengths), possibly
keeping into account the presence in our computers of various memory
levels, each with its technological and performance
characteristics~\cite{cpu}. The index design problem therefore turns
out to be more challenging than ever before.

\medskip
{\em The American Heritage Dictionary} (2000, fourth edition) defines
{\em index} as follows: pl. {\bf (in $\cdot$ dex $\cdot$ es)} or {\bf
(in $\cdot$ di $\cdot$ ces)} `` {\bf 1.} Something that serves to
guide, point out, or otherwise facilitate reference, especially: {\bf
a.} An alphabetized list of names, places, and subjects treated in a
printed work, giving the page or pages on which each item is
mentioned.  {\bf b.} A thumb index.  {\bf c.} Any table, file, or
catalog. [...]''

Some definitions proposed by experts are ``The most important of the
tools for information retrieval is the {\em index}---a collection of
terms with pointers to places where information about documents can be
found''~\cite{manber99}; ``{\em indexing} is building a data structure
that will allow quick searching of the text''~\cite{baeza99}; or ``the
act of assigning index terms to documents which are the objects to be
retrieved''~\cite{korfhage97}.

From our point of view an \emph{index} is a persistent data structure
that allows at query time to focus the search for a user-provided {\em
string} (or a set of them) on a very small portion of the indexed data
collection, namely the locations at which the queried string(s)
occur. Of course the index is just one of the tools needed to fully
solve a user query, so as the retrieval of the queried string
locations is just the first step of what is called the ``query
answering process''.  Information retrieval (IR) models, ranking
algorithms, query languages and operations, user-feedback models and
interfaces, and so on, all of them constitute the rest of this
complicated process and are beyond the scope of this survey.
Hereafter we will concentrate our attention onto the challenging
problems concerned with the design of efficient and effective indexing
data structures, the basic block upon which every IR system is
built. We then refer the reader interested into those other
interesting topics to the vast literature, browsing from
e.g.~\cite{baeza92,lesk,raghavan:97,baeza99,wmb99}.

\medskip
{\bf The right step into the text-indexing field.} The publications
regarding indexing techniques and methodologies are a common outcome
of database and algorithmic research. Their number is ever growing so
that citing all of them is a task doomed to fail. This fact is
contributing to make the evaluation of the novelty, impact and
usefulness of the plethora of recent index proposals more and more
difficult. Hence to approach from the correct angle the huge field of
{\em text indexing}, we first need a clear framework for development,
presentation and comparison of indexing schemes~\cite{guidelines}. The
lack of this framework has lead some researchers to underestimate the
features of known indexes, disregard important criteria or make
simplifying assumptions which have lead them to unrealistic and/or
distort results.

The design of a new index passes through the evaluation of many
criteria, not just its description and some toy experiments. We need
at a minimum to consider overall speed, disk and memory space
requirements, CPU time and measures of disk traffic (such as number of
seeks and volume of data transferred), and ease of index
construction. In a dynamic setting we should also consider index
maintenance in the presence of addition, modification and deletion of
documents/records; and implications for concurrency, transactions and
recoverability. Also of interest for both static and dynamic data
collections are applicability, extensibility and scalability. Indeed
no indexing scheme is all-powerful, different indexes support
different classes of queries and manage different kinds of data, so
that they may turn out to be useful in different application
contexts. As a consequence there is no one single winner among the
indexing data structures nowadays available, each one has its own
positive features and drawbacks, and we must know all of their fine
details in order to make the right choice when implementing an
effective and efficient search engine or IR system.

In what follows we therefore go into the main aspects which influence
the design of an indexing data structure thus providing an overall
view of the text indexing field; we introduce the arguments which will
be detailed in the next sections, and we briefly comment on some
recent topics of research that will be fully addressed at the end of
each of these subsequent sections.

\medskip
{\bf The first key issue: The I/O subsystem.}  The large amount of
textual information currently available in electronic form requires to
store it into external storage devices, like (multiple) disks and
cdroms. Although these mechanical devices provide a large amount of
space at low cost, their access time is more than $10^5$ times slower
than the time to access the internal memory of
computers~\cite{IEEEComputerIO}. This gap is currently widening with
the impressive technological progresses on circuit design
technology. Ongoing research on the engineering side is therefore
trying to improve the input/output subsystem by introducing some
hardware mechanisms such as disk arrays, disk caches,
etc.. Nevertheless the improvement achievable by means of a {\em
proper arrangement of data} and a properly {\em structured algorithmic
computation} on disk devices abundantly surpasses the best expected
technology advancements~\cite{Vitter}.

Larger datasets can stress the need for {\em locality of reference} in
that they may reduce the chance of sequential (cheap) disk accesses to
the same block or cylinder; they may increase the data fetch costs
(which are typically linear in the dataset size); and they may even
affect the proportion of documents/records that answer to a user
query. In this situation a na\"{\i}ve index might incur the so called
{\em I/O-bottleneck}, that is, its update and query operations might
spend most of the time in transferring data to/from the disk with a
consequent sensible slowdown of their performance. As a result, the
{\em index scalability} and the {\em asymptotic analysis} of index
performance, orchestrated with the {\em disk consciousness} of index
design, are nowadays hot and challenging research topics which have
shown to induce a positive effect not limited just to mechanical
storage devices, but also to all other memory levels (L1 and L2
caches, internal memory, etc.).

To design and carefully analyze the scalability and query performance
of an index we need a computational model that abstracts in a
reasonable way the {\em I/O-subsystem}.  Accurate disk models are
complex~\cite{Ruemmler-Wilkes}, and it is virtually impossible to
exploit all the fine points of disk characteristics systematically,
either in practice or for algorithmic design. In order to capture in
an easy, yet significant, way the differences between the internal
(electronic) memory and the external (mechanical) disk, we adopt the
{\em external memory model} proposed in~\cite{Vitter}. Here a computer
is abstracted to consist of a two-level memory: a fast and small
internal memory, of size $M$, and a slow and arbitrarily large
external memory, called {\em disk}. Data between the internal memory
and the disk are transfered in blocks of size $B$ (called {\em disk
pages}).  Since disk accesses are the dominating factor in the running
time of many algorithms, the asymptotic performance of the algorithms
is evaluated by counting the total number of disk accesses performed
during the computation.  This is a workable approximation for
algorithm design, and we will use it to evaluate the performance of
query and update algorithms. However there are situations, like in the
construction of indexing data structures (Sections~\ref{word:caching}
and~\ref{full:construction}), in which this accounting scheme does not
accurately predict the running time of algorithms on real machines
because it does not take into account some important specialties of
disk systems~\cite{quantum}. Namely, disk access costs have mainly two
components: the time to fetch the first bit of requested data (seek
time) and the time required to transmit the requested data (transfer
rate). Transfer rates are more or less stable but seek times are
highly variable.  It is thus well known that accessing one page from
the disk in most cases decreases the cost of accessing the page
succeeding it, so that ``bulk'' I/Os are less expensive per page than
``random'' I/Os.  This difference becomes much more prominent if we
also consider the reading-ahead/buffering/caching optimizations which
are common in current disks and operating systems.  To deal with these
specialties and avoid the introduction of many new parameters, we will
sometime refer to the simple accounting scheme introduced
in~\cite{FFM98}: a {\em bulk I/O} is the reading/writing of a
contiguous sequence of $cM/B$ disk pages, where $c$ is a proper
constant; a {\em random} I/O is any single disk-page access which is
not part of a bulk I/O.

In summary the performance of the algorithms designed to build,
process or query an indexing data structure is therefore evaluated by
measuring: (a)~the number of random I/Os, and possibly the bulk I/Os,
(b)~the internal running time (CPU~time), (c)~the number of disk pages
occupied by the indexing data structure and the working space of the
query, update and construction algorithms.

\medskip
{\bf The second key issue: types of queries and indexed data.}  Up to
now we have talked about indexing data structures without specifying
the type of queries that an index should be able to support as well no
attention has been devoted to the type of data an index is called to
manage. These issues have a surprising impact on the design complexity
and space occupancy of the index, and will be strictly interrelated in
the discussion below.

There are two main approaches to index design:~\emph{word-based}
indexes and \emph{full-text} indexes.  Word-based indexes are designed
to work on linguistic texts, or on documents where a {\em
tokenization} into {\em words} may be devised.  Their main idea is to
store the occurrences of each word (token) in a table that is indexed
via a hashing function or a tree structure (they are usually called
{\em inverted files} or {\em indexes}). To reduce the size of the
table, common words are either not indexed (e.g.  the, at, a) or the
index is later compressed. The advantage of this approach is to
support very fast word (or prefix-word) queries and to allow at
reasonable speed some complex searches like regular expression or
approximate matches; while two weaknesses are the impossibility in
dealing with \emph{non-tokenizable} texts, like genomic sequences, and
the slowness in supporting arbitrary {\em substring}
queries. Section~\ref{word-based} will be devoted to the discussion of
word-based indexes and some recent advancements on their
implementation, compression and supported operations. Particular
attention will be devoted to the techniques used to compress the
inverted index or the input data collection, and to the algorithms
adopted for implementing more complex queries.

Full-text indexes have been designed to overcome the limitations above
by dealing with arbitrary texts and general queries, at the cost of an
increase in the additional space occupied by the underlying data
structure.  Examples of such indexes are: suffix trees~\cite{McC76},
suffix arrays~\cite{MM93} and String B-trees~\cite{FG:99}.  They have
been successfully applied to fundamental string-matching problems as
well to text compression~\cite{bw}, analysis of genetic
sequences~\cite{Gus97}, optimization of Xpath queries on XML
documents~\cite{fabric,lore1,aboulnaga} and to the indexing of special
linguistic texts~\cite{chinese}.  General full-text indexes are
therefore the natural choice to perform fast complex searches without
any restrictions on the query sequences and on the format of the
indexed data; however, a reader should always keep in mind that these
indexes are usually more space demanding than their word-based
counterparts~\cite{Kurtz-st,CM96} (cfr. opportunistic
indexes~\cite{FM:00} below).  Section~\ref{full-text} will be devoted
to a deep discussion on full-text indexes, posing particular attention
to the String B-tree data structure and its engineering. In particular
we will introduce some novel algorithmic and data structural solutions
which are not confined to this specific data structure. Attention will
be devoted to the challenging, yet difficult, problem of the
construction of a full-text index both from a theoretical and a
practical perspective. We will show that this problem is related to
the more general problem of {\em string sorting}, and then discuss the
known results and a novel randomized algorithm which may have
practical utility and whose technical details may have an independent
interest.

\medskip
{\bf The third key issue: the space vs. time trade-off.} The
discussion on the two indexing approaches above has pointed out an
interesting trade-off: space occupancy vs. flexibility and efficiency
of the supported queries. It indeed seems that in order to support
substring queries, and deal with arbitrary data collections, we do
need to incur in an additional space overhead required by the more
complicated structure of the full-text indexes. Some authors argue
that this extra-space occupancy is a false problem because of the
continued decline in the cost of external storage devices. However the
impact of space reduction goes far beyond the intuitive memory saving,
because it may induce a better utilization of (the fast) cache and
(the electronic) internal memory levels, may virtually expand the disk
bandwidth and significantly reduce the (mechanical) seek time of disk
systems. Hence data compression is an attractive choice, if not
mandatory, not only for storage saving but also for its favorable
impact on algorithmic performance. This is very well known in
algorithmics~\cite{Knuth:1998:SS} and engineering~\cite{ibm-mxt}: IBM
has recently delivered the MXT Technology (Memory eXpansion
Technology) for its x330 eServers which consists in a memory chip that
compresses/decompresses data on cache writebacks/misses thus yielding
a factor of expansion two on memory size with just a slightly larger
cost.  It is not surprising, therefore, that we are witnessing in the
algorithmic field an upsurging interest for designing {\em succinct}
(or {\em implicit}) data structures (see
e.g.~\cite{Brodnik:1999,Munro97,Munro98,Munro99,GV:00,sada00,sada-soda02})
that try to reduce as much as possible the auxiliary information kept
for indexing purposes without introducing any significant slowdown in
the operations supported.

Such a research trend has lead to some surprising results on the
design of {\em compressed full-text indexes}~\cite{FM:00} whose impact
goes beyond the text-indexing field. These results lie at the crossing
of three distinct research fields--- compression, algorithmics,
databases--- and orchestrate together their latest achievements, thus
showing once more that the design of an indexing data structure is
nowadays an interdisciplinary task. In Section~\ref{trade-off} we will
briefly overview this issue by introducing the concept of {\em
opportunistic index}: a data structure that tries to take advantage of
the compressibility of the input data to reduce its overall space
occupancy. This index encapsulates both the compressed data and the
indexing information in a space which is proportional to the entropy
of the indexed collection, thus resulting optimal in an
information-content sense. Yet these results are mainly theoretical in
their flavor and open to significant improvements with respect to
their I/O performance. Some of them have been implemented and tested
in~\cite{is01,FM:01} showing that these data structures use roughly
the same space required by traditional compressors---such as \gzip\
and \bzip~\cite{bzip2_home}--- but with added functionalities: they
allow to retrieve the occurrences of an arbitrary substring within
texts of several megabytes in a few milliseconds. These experiments
show a promising line of research and suggest the design of a new
family of text retrieval tools which will be discussed at the end of
Section~\ref{trade-off}.

\medskip
{\bf The fourth key issue: String transactions and index caching.}
Not only is string data proliferating, but datastores increasingly
handle large number of {\em string transactions} that add, delete,
modify or search strings. As a result, the problem of managing massive
string data under large number of transactions is emerging as a
fundamental challenge. Traditionally, string algorithms focus on
supporting each of these operations {\em individually} in the most
efficient manner in the worst case. There is however an ever
increasing need for indexes that are efficient on an entire sequence
of string transactions, by possibly adapting themselves to
time-varying distribution of the queries and to the repetitiveness
present in the query sequence both at string or prefix level. Indeed
it is well known that some user queries are frequently issued in some
time intervals~\cite{two-level-cache} or some search engines improve
their precision by expanding the query terms with some of their
morphological variations (e.g. synonyms, plurals,
etc.)~\cite{baeza99}. Consequently, in the spirit of {\em amortized
analysis}~\cite{Sleator-Tarjan}, we would like to design indexing data
structures that are competitive (optimal) over the entire sequence of
string operations.  This challenging issue has been addressed at the
heuristic level in the context of word-based
indexes~\cite{two-level-cache,cache-brown,cache-markatos,cache-meira,cache-sriva};
but it has been unfortunately disregarded when designing and analyzing
full-text indexes. Here the problem is particularly difficult because:
(1)~a string may be so long to do not fit in one single disk page or
even be contained into internal memory, (2)~each string comparison may
need many disk accesses if executed in a brute-force manner, and
(3)~the distribution of the string queries may be unknown or vary over
the time.  A first, preliminary, contribution in this setting has been
achieved in~\cite{ciriani-et-al} where a self-adjusting and
external-memory variant of the skip-list data structure~\cite{Pugh90}
has been presented. By properly orchestrating the caching of this data
structure, the caching of some query-string prefixes and the effective
management of string items, the authors prove an {\em external-memory
version for strings} of the famous Static Optimality
Theorem~\cite{Sleator-Tarjan}. This introduces a new framework for
designing and analyzing full-text indexing data structures and
string-matching algorithms, where a {\em stream of user queries} is
issued by an unknown source and caching effects must then be exploited
and accounted for when analyzing the query operations. In the next
sections we will address the caching issue both for word-based and
full-text indexing schemes, pointing out some interesting research
topics which deserve a deeper investigation.

\bigskip
The moral that we would like to convey to the reader is that the text
indexing field has grown at a such complicated stage that various
issues come into play when studying it: data structure design,
database principles, compression techniques, architectural
considerations, cache and prefetching policies. The expertise nowadays
required to design a good index is therefore transversal to many
algorithmic fields and much more study on the orchestration of known,
or novel, techniques is needed to make progress in this fascinating
topic. The rest of the survey is therefore devoted to illustrate the
key ideas which should constitute, in our opinion, the current
background of every index-designer. The guiding principles of our
discussion will be the four key issues above; they will guide the
description of the positive features and drawbacks of known indexing
schemes as well the investigation of research issues and open
problems. A vast, but obviously not complete, literature will
accompany our discussion and should be the reference where an eager
reader may find further technical details and research hints.

%%%%%%%%%%%%%%%%%% WORD-BASED %%%%%%%%%%%%%%%%%%%%%%%%%%

\section{On the word-based indexes}
\label{word-based}

There are three main approaches to design a word-based index: inverted
indexes, signature files and
bitmaps~\cite{wmb99,baeza99,baeza-chapter,Faloutsos:1985:AMT}. The
inverted index--- also known as {\em inverted file}, {\em posting
file}, or in normal English usage as {\em concordance}--- is doubtless
the simplest and most popular technique for indexing large text
databases storing natural-language documents. The other two mechanisms
are usually adopted in certain applications even if, recently, they
have been mostly abandoned in favor of inverted indexes because some
extensive experimental results~\cite{Zobel:1998:IFV} have shown that:
{\em Inverted indexes offer better performance than signature files
and bitmaps, in terms of both size of index and speed of query
handling}~\cite{wmb99}. As a consequence, the emphasis of this section
is on inverted indexing; a reader interested into signature files
and/or bitmaps may start browsing from~\cite{wmb99,baeza99} and have a
look to some more recent, correlated and stimulating results
in~\cite{bloom,mitzenmacher}.

An inverted index is typically composed of two parts: the {\em
lexicon}, also called the {\em vocabulary}, containing all the
distinct words of the text collection; and the {\em inverted list},
also called the {\em posting list}, storing for each vocabulary term a
list of all text positions in which that term occurs. The vocabulary
therefore supports a mapping from words to their corresponding
inverted lists and in its simplest form is a list of strings and disk
addresses. The search for a single word in an inverted index consists
of two main phases: it first locates the word in the vocabulary and
then retrieves its list of text positions. The search for a phrase or
a proximity pattern (where the words must appear consecutively or
close to each other, respectively) consists of three main phases: each
word is searched separately, their posting lists are then retrieved
and finally intersected, taking care of consecutiveness or closeness
of word positions in the text.

It is apparent that the inverted index is a simple and natural
indexing scheme, and this has obviously contributed to its spread
among the IR systems. Starting from this simple theme, researchers
indulged theirs whims by proposing numerous variations and
improvements. The main aspect which has been investigated is the {\em
compression} of the vocabulary and of the inverted lists. In both
cases we are faced with some challenging problems.

Since the vocabulary is a textual file any classical compression
technique might be used, provided that subsequent pattern searches can
be executed efficiently. Since the inverted lists are constituted by
numbers any {\em variable length encoding} of integers might be used,
provided that subsequent sequential decodings can be executed
efficiently. Of course, any choice in vocabulary or inverted lists
implementation influences both the processing speed of queries and the
overall space occupied by the inverted index. We proceed then to
comment each of these points below, referring the reader interested
into their fine details to the cited literature.

The vocabulary is the basic block of the inverted index and its
``content'' constraints the type of queries that a user can
issue. Actually the index designer is free to decide {\em what a word}
is, and which are the {\em representative words} to be included into
the vocabulary. One simple possibility is to take each of the words
that appear in the document and declare them verbatim to be vocabulary
terms. This tends both to enlarge the vocabulary, i.e. the number of
distinct terms that appear into it, and increase the number of
document/position identifiers that must be stored in the posting
lists. Having a large vocabulary not only affects the storage space
requirements of the index but can also make it harder to use since
there are more potential query terms that must be considered when
formulating a query. For this reason it is common to transform each
word in some {\em normal form} before being included in the
vocabulary. The two classical approaches are {\em case folding}, the
conversion of all uppercase letters to their lowercase equivalents (or
vice versa), and {\em stemming}, the reduction of each word to its
morphological root by removing suffixes or other modifiers. It is
evident that both approaches present advantages (vocabulary
compression) and disadvantages (extraneous material can be retrieved
at query time) which should be taken into account when designing an IR
system. Another common transformation consists of omitting the so
called {\em stop words} from the indexing process (e.g., a, the, in):
They are words which occur too often or carry such small information
content that their use in a query would be unlikely to eliminate any
documents. In the literature there has been a big debate on the
usefulness of removing or keeping the stop words. Recent progresses on
the compaction of the inverted lists have shown that the space
overhead induced by those words is not significant, and is abundantly
payed for by the simplification in the indexing process and by the
increased flexibility of the resulting index.

The size of the vocabulary deserves a particular attention. It is
intuitive that it should be small, but more insight on its cardinality
and structure must be acquired in order to go into more complex
considerations regarding its compression and querying.  An empirical
law widely accepted in IR is the Heaps' Law~\cite{Hea78}, which states
that the vocabulary of a text of $n$ words is of size $V=O(n^\beta)$,
where $\beta$ is a small positive constant depending on the text. As
shown in~\cite{igrep}, $\beta$ is practically between $0.4$ and $0.6$
so the vocabulary needs space proportional to the square root of the
indexed data. Hence for large data collections the overhead of storing
the vocabulary, even in its extended form, is minimal.  Classical
implementations of a set of words via {\em hash tables} and {\em trie}
structures seem appropriate for exact word or prefix word queries. As
soon as the user aims for more complicated queries, like approximate
or regular-expression searches, it is preferable to keep the
vocabulary in its plain form as a {\em vector of words} and then
answer a user query via one of the powerful scan-based string-matching
algorithms currently known~\cite{navarro-survey}. The increase in
query time is payed for by the more complicated queries the index is
able to support.

As we observed in the Introduction, space saving is intimately related
to time optimization in a hierarchical memory system, so that it turns
out to be natural to ask ourselves if, and how, compression can help
in vocabulary storage and searching. From one hand, vocabulary
compression might seem useless because of its small size; but from the
other hand, any improvement in the vocabulary search-phase it is
appealing because the vocabulary is examined at each query on all of
its constituting terms. Numerous scientific
results~\cite{ABF96,Man94b,GKPR96,GKPR96b,tm97:acsc,FT98,Moura98,KTSA99,NR99b,STSA99,AD99,MNZB00,navarro-regex,approx-lz}
have recently shown how to compress a textual file and perform exact
or approximate searches directly on the compressed text without
passing through its whole decompression. This approach may be
obviously applied to vocabularies thus introducing two immediate
improvements: it squeezes them to an extension that can be easily kept
into internal memory even for large data collections; it reduces the
amount of data examined during the query phase, and it fully exploits
the processing speed of current processors with respect to the
bandwidth and access time of internal memories, thus impacting
fruitfully onto the overall query performance. Experiments have shown
a speed up of a factor about two in query processing and a reduction
of more than a factor three in space occupancy. Nonetheless the whole
scanning of the compressed dictionary is afforded, so that some room
for query time improvement is still possible. We will be back on this
issue in Section~\ref{trade-off}.

Most of the space usage of inverted indexes is devoted to the storage
of the inverted lists; a proper implementation for them thus becomes
urgent in order to make such an approach competitive against the other
word-based indexing methods: signature files and
bitmaps~\cite{wmb99,Zobel:1998:IFV}. A large research effort has been
therefore devoted to effectively compress the inverted lists still
guaranteeing a fast sequential access to their contents. Three
different types of compaction approaches have been proposed in the
literature, distinguished according to the {\em accuracy} to which
the inverted lists identify the location of a vocabulary term, usually
called {\em granularity} of the index. A {\em coarse-grained} index
identifies only the documents where a term occurs; an index of {\em
moderate-grain} partitions the texts into blocks and stores the block
numbers where a term occurs; a {\em fine-grained} index returns
instead a sentence, a term number, or even the character position of
every term in the text. Coarse indexes require less storage (less than
25\% of the collection size), but during the query phase parts of the
text must be scanned in order to find the exact locations of the query
terms; also, with a coarse index multi-term queries are likely to give
rise to {\em insignificant matches}, because the query terms might
appear in the same document but far from each other. At the other
extreme, a word-level indexing enables queries involving adjacency and
proximity to be answered quickly because the desired relationship can
be checked without accessing the text. However, adding precise
locational information expands the index of at least a factor of two
or three, compared with a document-level indexing since there are more
pointers in the index and each one requires more bits of storage. In
this case the inverted lists take nearly 60\% of the collection
size. Unless a significant fraction of the queries are expected to be
proximity-based, or ``snippets'' containing text portions where the
query terms occur must be efficiently visualized, then it is
preferable to choose a document-level granularity; proximity and
phrase-based queries as well snippet extraction can then be handled by
a post-retrieval scan.

In all those cases the size of the resulting index can be further
squeezed down by adopting a compression approach which is orthogonal
to the previous ones. The key idea is that each inverted list can be
sorted in increasing order, and therefore the {\em gaps} between
consecutive positions can be stored instead of their absolute
values. Here can be used compression techniques for small integers. As
the gaps for longer lists are smaller, longer lists can be compressed
better and thus stop words can be kept without introducing a
significant overhead in the overall index space. A number of suitable
codes are described in detail in~\cite{wmb99}, more experiments are
reported in~\cite{wz99}. Golomb codes are suggested as the best ones
in many situations, e.g. TREC collection, especially when the integers
are distributed according to a geometric law.  Our experience however
suggests to use a simpler, yet effective, coding scheme which is
called {\em continuation bit} and is currently adopted in Altavista
and Google search engines for storing compactly their inverted
lists. This coding scheme yields a byte-aligned and compact
representation of an integer $x$ as follows. First, the binary
representation of $x$ is partitioned into groups of 7 bits each,
possibly appending zeros to its beginning; then, one bit is appended
to the front of each group setting it to one for the first group and
to zero for the other groups; finally, the resulting sequence of 8-bit
groups is allocated to a contiguous sequence of bytes. The
byte-aligning ensures fast decoding/encoding operations, whereas the
tagging of the first bit of every byte ensures the fast detection of
codeword beginnings. For an integer $x$, this representation needs
$\lfloor \log_2 x + 1 \rfloor /7$ bytes; experiments show that its
overhead wrt Golomb codes is small, but the Continuation bit scheme is
by far much faster in decoding thus resulting the natural choice
whenever the space issue is not a main concern.  If a further space
overhead is allowed and queries have to be speeded up, other integer
coding approaches do exist. Among the others we cite the {\em
frequency sorted} index organization of~\cite{persin}, which sorts the
posting lists in decreasing order of frequency to facilitate the
immediate retrieval of relevant occurrences, and the {\em blocked
index} of~\cite{ahn} which computes the gaps with respect to some
equally-sampled pivots to avoid the decoding of some parts of the
inverted lists during their intersection at query time.

There is another approach to index compression which encompasses all
the others because it can be seen as their generalization. It is
called {\em block-addressing index} and was introduced in a system
called {\em Glimpse} some years ago~\cite{glimpse}. The renewed
interest toward it is due to some recent results~\cite{NMN00,FM:00}
which have shed new light on its structure and opened the door to
further improvements. In this indexing scheme, the whole text
collection is divided into {\em blocks} of fixed size; these blocks
may span many documents, be part of a document, or overlap document
boundaries. The index stores only the block numbers where each
vocabulary term appears. This introduces two space savings: multiple
occurrences of a vocabulary term in a block are represented only once,
and few bits are needed to encode a block number. Since there are
normally much less blocks than documents, the space occupied by the
index is very small and can be tuned according to the user needs. On
the other hand, the index may by used just as a device to identify
some {\em candidate blocks} which may contain a query-sting
occurrence. As a result a post-processing phase is needed to filter
out the candidate blocks which actually do not contain a match
(e.g. the block spans two documents and the query terms are spread in
both of them). As in the document-level indexing scheme,
block-addressing requires very little space, close to 5\% of the
collection size~\cite{glimpse}, but its query performance is modest
because of the postprocessing step and critically depends on the {\em
block size}. Actually by varying the block size we can make the
block-addressing scheme to range from coarse-grained to fine-grained
indexing. The smaller the block size, the closer to a word-level index
we are, the larger is the index but the faster is the query
processing. On the other extreme, the larger is the block size, the
smaller is the space occupancy but the larger is the query
time. Finding a good trade-off between these two quantities is then a
matter of user needs; the analysis we conduct below is based on some
reasonable assumptions on the distribution of the vocabulary terms and
the linguistic structure of the
documents~\cite{CIKM97,CIKM97_jou}. This allows us to argue about some
positive features of the block-addressing scheme.

The Heaps' law, introduced above, gives a bound on the vocabulary
size. Another useful law related to the vocabulary is the Zipf's
Law~\cite{zipf} which states that, in a text of $n$ terms, the $i$th
most frequent term appears $n/(i^\theta z)$ times, where $\theta$ is a
constant that depends on the data collection (typical~\cite{Harman}
experimental values are in $[1.7,2.0]$) and $z$ is a normalization
factor.  Given this model, it has been shown in~\cite{CIKM97_jou} that
the block-addressing scheme may achieve $O(n^{0.85})$ space and query
time complexity; notice that both complexities are sublinear in the
data size. 

Apart from this analytical calculations, it is apparent that speeding
up the postprocessing step (i.e. the scanning of candidate blocks)
would impact on the query performance of the index. This was the
starting point of the fascinating paper~\cite{NMN00} which
investigated how to combine in a single scheme: index compression,
block addressing and sequential search on compressed text. In this
paper the specialized compression technique of~\cite{MNZB00} is
adopted to squeeze each text block in less than 25\% of its original
size, and perform direct searching on the compressed candidate blocks
without passing through their whole decompression. The specialty of
this compression technique is that it is a variant of the Huffman's
algorithm with {\em byte-aligned} and {\em tagged} codewords. Its
basic idea is to build a Huffman tree with fan-out 128, so that the
binary codewords have length a multiple of 7 bits. Then these
codewords are partitioned into groups of 7 bits; to each group is
appended a bit that is set to 1 for the first group and to 0 for the
others; finally, each 8-bit group is allocated to a byte. The
resulting codewords have many nice properties: (1)~they are
byte-aligned, hence their decoding is fast and requires very few
shift/masking operations; (2)~they are tagged, hence the beginning of
each codeword can be easily identified; (3)~they allow exact
pattern-matching directly over the compressed block, because {\em no
tagged codeword can overlap more then two tagged codewords}; (4)~they
allow the search for more complex patterns directly on the compressed
blocks~\cite{MNZB00,NMN00}. The overall result is an improvement of a
factor about 3 over well known tools like Agrep~\cite{MW92} and
Cgrep~\cite{MNZB00}, which operate on uncompressed blocks. If we add
to these interesting features the fact that the symbol table of this
Huffman's variant is actually the vocabulary of the indexed
collection, then we may conclude that this approach couples perfectly
well with the inverted-index scheme.

\begin{figure}[t]
\centerline{\psfig{file=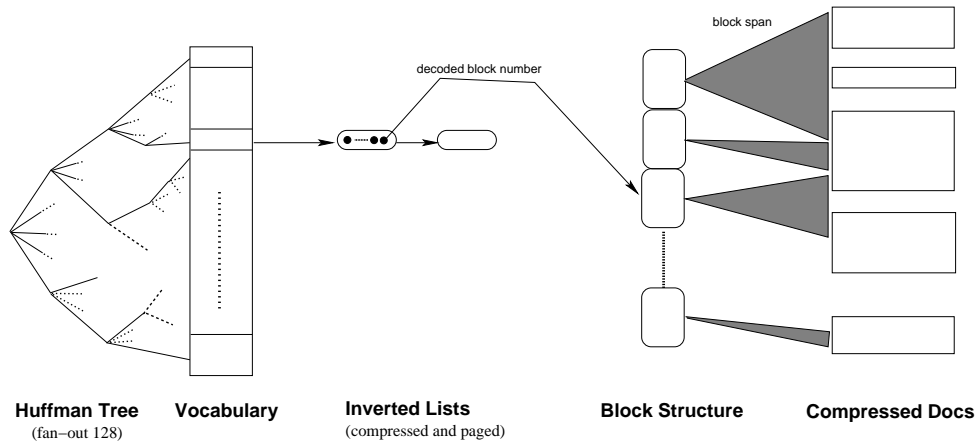,width=13cm}} 
\caption{\em The highlevel structure of the block-addressing
scheme. \label{fig:block-scheme}}
\end{figure}

Figure~\ref{fig:block-scheme} provides a pictorial summary of the
block-addressing structure. We will be back on this approach in
Section~\ref{trade-off} where we discuss and analyze a novel
compressed index for the candidate blocks which has opened the door to
further improvements.

\subsection{Constructing an inverted index}
\label{word:caching}
 
This journey among the inverted index variations and results has
highlighted some of their positive features as well their
drawbacks. It is clear that the structure of the inverted index is
suitable to be mapped in a two-level memory system, like the
disk/memory case. The vocabulary can be kept in internal memory, it is
usually small and random accesses must be performed on its terms in
order to answer the user queries; the inverted lists can be allocated
on disk each in a contiguous sequence of disk pages, thus fully
exploiting the prefetching/caching capabilities of current disks
during the subsequent gap-decoding operations. In this case the
performance of current processors is sufficient to make transparent
the decoding cost with respect to the one incurred for fetching the
compressed lists from the disk.

There is however another issue which has been not addressed in the
previous sections and offers some challenging problems to be deal
with. It concerns with the construction of the inverted lists.  Here,
the I/O-bottleneck can play a crucial role, and a na\"{\i}ve algorithm
might be unable to build the index even for collections of moderate
size. The use of in-memory data structures of size larger than the
actual internal memory and the non sequential access to them, might
experience a so high paging activity of the system to require one I/O
per operation~!  Efficient methods have been presented in the
literature~\cite{mb95,wmb99} to allow a more economical index
construction. From an high-level point of view, they follow an
algorithmic scheme which recalls to our mind the multiway mergesort
algorithm; however, the specialties of the problem make compression a
key tool to reduce the volume of processed data and constraint to
reorganize the operations in order to make use of sequential
disk-based processing. For the sake of completeness we sketch here an
algorithm that has been used to build an inverted index over a
multi-gigabyte collection of texts within few tens of megabytes of
internal memory and only a small amount of extra disk space. The
algorithm will be detailed for the case of a document-level indexing
scheme, other extensions are possible and left to the reader as an
exercise. The basis of the method is a process that creates a file of
pairs $\langle d,t\rangle$, where $d$ is a document number and $t$ is
a term number. Initially the file is ordered by increasing $d$, then
the file is reordered by increasing $t$ using an in-place multi-way
external mergesort. This sorting phase is then followed by an in-place
permutation of the disk pages that collectively constitute the
inverted lists in order to store each of them into a consecutive
sequence of disk pages.

In detail, the collection is read in document order and parsed into
terms, which will form the vocabulary of the inverted index. A bounded
amount of internal memory is set aside as a working buffer. Pairs
$\langle d,t\rangle$ are collected into the buffer until it is full;
after that, it is sorted according to the term numbers and a run of
disk pages is written to disk in a compressed format (padding is used
to get disk-page alignment). Once all the collection has been
processed, the resultant runs are combined via a multiway merge: Just
one block of each run is resident in memory at any given time, and so
the memory requirement is modest. As the merge proceeds, output blocks
are produced and written back to disk (properly compressed) to any
available slot. Notice that there will be always one slot available
because the reading (merging) process frees the block slots at a
faster rate than the blocks consumed by the writing process. Once all
the runs have been exhausted, the index is complete, but the inverted
lists are spread over the disk so that locality of reference is absent
and this would slowdown the subsequent query operations. An in-place
permutation is then used to reorder the blocks in order to allocate
each inverted list into a contiguous sequence of disk pages. This step
is disk-intensive, but usually executed for a short amount of time. At
the end a further pass on the lists can be executed to ``refine''
their compression; any now-unused space at the end of the file can be
released. Experimental results~\cite{wmb99,NMN00} have shown that the
amount of internal memory dedicated to the sorting process impacts a
lot, as expected, on the final time complexity. Just to have an idea,
a 5~Gb collection can be inverted using an internal memory space which
is just the one required for the vocabulary, and a disk space which is
about 10\% more than the final inverted lists, at an overall rate of
about 300~Mb of text per hour~\cite{wmb99}. If more internal memory is
reserved for the sorting process, then we can achieve an overall rate
of about 1~Gb of text per hour~\cite{NMN00}.

\subsection{Some open problems and future research directions}
\label{open-wordbased}

We conclude this section by addressing some other interesting
questions which, we think, deserve some attention and further
investigation.  First, we point out one challenging feature of the
block-addressing scheme which has been not yet fully exploited: the
vocabulary allows to turn approximate or complex pattern queries on
the text collection into an exact search for, possibly many,
vocabulary terms on the candidate blocks (i.e. the vocabulary terms
matching the complex user query). This feature has been deployed in
the solutions presented in~\cite{MNZB00,NMN00} to speed up the whole
scanning of the compressed candidate blocks. We point out here a
different perspective which may help in further improving the
postprocessing phase. Indeed we might build a succinct index that
supports {\em just} exact pattern searches on each compressed blocks,
and then use it in combination with the block-addressing scheme to
support arbitrarily complex pattern searches. This index would gain
powerful queries, reduced space occupancy and, more importantly, a
faster search operation because the cost of a candidate-block
searching could be $o(b)$. This would impact onto the overall index
design and performance. A proposal in this direction has been pursued
in~\cite{FM:00}, where it has been shown that this novel approach
achieves {\em both space overhead and query time sublinear in the data
collection size independently of the block size $b$}. Conversely,
inverted indices achieve only the second goal~\cite{wmb99}, and
classical block-addressing schemes achieve both goals but under some
restrictive conditions on the value of $b$~\cite{CIKM97_jou}.

Another interesting topic of research concerns with the design of
indices and methods for supporting faster vocabulary searches on
complex pattern queries. Hashing or trie structures are well suited to
implement (prefix)word queries but they actually fail in supporting
suffix, substring or approximate word searches. In these cases the
common approach consists of scanning the whole vocabulary, thus
incurring in a performance slowdown that prevents its use in search
engines aiming for a high throughput. Filtering
methods~\cite{navarro-survey} as well novel metric
indexes~\cite{chavet-navarro} might possibly help in this respect but
simple, yet effective, data structures with provable query bounds are
still to be designed.

We have observed that the block-addressing scheme and gap-coding
methods are the most effective tools to squeeze the posting lists in a
reduced space. A gap-coding algorithm achieves the best compression
ratio if most of the differences are very small. Several
authors~\cite{BKR94,BKR95,MofStu96} have noticed that this occurs when
the document numbers in each posting list have high locality, and
hence they designed methods to passively exploit this locality
whenever present in the posting lists. A different approach to this
problem has been undertaken recently in~\cite{blandford-blelloch}
where the authors suggest to permute the document numbers in order to
actively create the locality in the individual posting lists. The
authors propose therefore a hierarchical clustering technique which is
applied on the document collection as a whole, using the cosine
measure as a basis of document similarity. The hierarchical clustering
tree is then traversed in preorder and numbers are assigned to the
documents as they are encountered. The authors argue that documents
that share many term lists should be close together in the tree, and
therefore be labeled with near numbers. This idea was tested on the
TREC-8 data (disks 4 and 5, excluding the Congressional Record), and
showed a space improvement of 14\%. Different similarity measures to
build the hierarchical tree, as well different clustering approaches
which possibly do not pass through the exploration of the complete
graph of all documents, constitute good avenues for research.

Another interesting issue is the exploitation of the large internal
memory currently available in our PCs to improve the query
performance. A small fraction of the internal memory is already used
at run time to maintain the vocabulary of the document terms and thus
to support fast word searches in response to a user query. It is
therefore natural to aim at using the rest of the internal memory to
{\em cache} parts of the inverted index or the last query answers, in
order to exploit the {\em reference and temporal locality} commonly
present in the query streams~\cite{web-locality,web-log} for achieving
improved query performance.  Due to the ubiquitous use of inverted
lists in current web search engines, and the ever increasing amount of
user queries issued per day, the design of caching methodologies
suitable for inverted-indexing schemes is becoming a hot topic of
research. Numerous papers have been recently published on this
subject, see
e.g.~\cite{two-level-cache,cache-brown,cache-markatos,cache-meira,cache-sriva},
which offer some challenging problems for further study: how the
interplay between the retrieval and ranking phase impacts on the
caching strategy, how the compression of inverted lists affects the
behavior of caching schemes, how to extend the caching ideas developed
for stand-alone machines to a distributed information retrieval
architecture~\cite{cache-meira,cache-molina}. We refer the reader to
the latest WWW, VLDB and SIGMOD/PODS conferences for keeping track of
this active research field.

On the software development side, there is much room for data
structural and algorithmic engineering as well code tuning and library
design. Here we would like to point out just one of the numerous
research directions which encompasses the interesting XML
language~\cite{xml-w3c}. XML is an extremely versatile markup
language, capable of labeling the information content of diverse data
sources including structured or semi-structured documents, relational
databases and object repositories. A query issued on XML documents
might exploit intelligently their structure to manage uniformly all
these kinds of data and to enrich the precision of the query
answers. Since XML was completed in early 1998 by the World Wide Web
Consortium~\cite{xml-w3c}, it has spread through science and industry,
thus becoming a de facto standard for the publication and interchange
of structured data over the Internet and amongst applications. The
turning point is that XML allows to represent the semantics of data in
a structured, documented, machine-readable form. This has lead some
researchers to talk about ``semantic Web'' in order to capture the
idea of having data on the Web defined and linked in a way that can be
used by machines not just for display (cfr. HTML), but for automation,
integration, reuse across various applications and, last but not
least, for performing ``semantic searches''. This is nowadays a vision
but a huge number of people all around the world are working to its
concretization.  One of the most tangible results of this effort is
the plethora of IR systems specialized today to work on XML
data~\cite{survey,bus,tox,tamino,xyleme,xyzfind,lore1,lorel,fabric,natix,xmlfs}. Various
approaches have been undertaken for their implementation but the most
promising for flexibility, space/time efficiency and complexity of the
supported queries is doubtless the one based on a ``native''
management of the XML documents via inverted
indexes~\cite{baeza-xml1,baeza-xml2}. Here the idea is to support
structured text queries by indexing (real or virtual) tags as distinct
terms and then answering the queries via complex combinations of
searches for words and tags. In this realm of solutions there is a
lack of a public, easily usable and customizable repository of
algorithms and data structures for indexing and querying XML
documents. We are currently working in this
direction~\cite{ferragina-mastroianni}: at the present time we have a
C library, called {\tt XCDE Library} ({\tt XCDE} stands for Xml
Compressed Document Engine) that provides a set of algorithms and data
structures for indexing and searching an XML document collection in
its ``native'' form.  The library offers various features:
state-of-the-art algorithms and data structures for text indexing,
compressed space occupancy, and novel succinct data structures for the
management of the hierarchical structure present into the XML
documents. Currently we are using the XCDE Library to implement a
search engine for a collection of Italian literary texts marked with
XML-TEI. The XCDE Library offers to a researcher the possibility to
investigate and experiment novel algorithmic solutions for indexing
and retrieval without being obliged to re-write from scratch all the
basic procedures which constitute the kernel of any classic IR system.

%%%%%%%%%%%%%%%%%%%%%% FULL-TEXT %%%%%%%%%%%%%%%%%%%%%%%%%%%%%%%%%%%

\section{On the full-text indexes}
\label{full-text}

The inverted-indexing scheme, as well any other word-based indexing
method, is well suited to manage text retrieval queries on linguistic
texts, namely texts composed in a natural language or properly
structured to allow the identification of ``terms'' that are the units
upon which the user queries will be formulated. Other assumptions are
usually made to ensure an effective use of this indexing method: the
text has to follow some statistical properties that ensure, for
example, small vocabulary size, short words, queries mostly concerning
with rare terms and aiming at the retrieval of parts of words or
entire phrases. Under these restrictions, which are nonetheless
satisfied in many practical user settings, the inverted indexes are
{\em the} choice since they provide efficient query performance, small
space usage, cheap construction time, and allow the easy
implementation of effective ranking techniques.

Full-text indexes, on the other hand, overcome the limitations of the
word-based indexes. They allow to manage any kind of data and support
complex queries that span arbitrary long parts of them; they allow to
draw statistics from the indexed data, as well make many kind of
complex text comparisons and investigations: detect pattern motifs,
auto-repetitions with and without errors, longest-repeated strings,
etc.. The full-text indexes may be clearly applied to classical
information retrieval, but they are less adeguate than inverted
indexes since their additional power comes at some cost: they are more
expensive to build and occupy significant more space. The real
interest in those indexing data structures is motivated by some
application settings where inverted indexes result unappropriate, or
even unusable: Building an inverted index on all the substrings of the
indexed data would need quadratic space~! The applications we have in
mind are: genomic databases (where the data collection consists of DNA
or protein sequences), intrusion detection (where the data are
sequences of events, log of accesses, along the time), oriental
languages (where word delimiters are not so clear), linguistic
analysis of the text statistics (where the texts are composed by words
but the queries require complex statistical elaborations to detect
plagiarism, for instance), Xpath queries in XML search engines (where
the indexed strings are paths into the hierarchical tree structure of
an XML document), and vocabulary implementations to support exact or
complex pattern searches (even the inverted indexes might benefit of
full-text indexes~!).

These fascinating properties and the powerful nature of full-text
indexes are the starting points of our discussion. To begin with we
need some notations and definitions.  

For the inverted indexes we defined as {\em index points} the block
numbers, word numbers or word starts in the indexed text. In the
context of full-text indexes an index point is, instead, any character
position or, classically, any position where a {\em text suffix} may
start. In the case of a text collection, an index point is an integer
pair $(j, i)$, where $i$ is the starting position of the suffix in the
$j$th text of the collection.  In most current applications, an index
point is represented using from $3$ to $6$ bytes, thus resulting
independent on the actual length of the pointed suffix, and characters
are encoded as {\em bit sequences}, thus allowing the uniform
management of arbitrary large alphabets.

Let $\Sigma$ be an arbitrarily large alphabet of characters, and let
$\#$ be a new character larger than any other alphabet character. We
denote by $lcp(P,Q)$ the longest common prefix length of two strings
$P$ and $Q$, by $max\_lcp(P,\S)$ the value $\max{\{ lcp(P,Q) : Q \in
\S \}}$, and by $\leq_L$ the lexicographic order between pair of
strings drawn from $\Sigma$. Finally, given a text $T[1,n]$, we denote
by $\suf(T)$ the lexicographically ordered set of all suffixes of text
$T$.

Given a pattern $P[1,p]$, we say that there is an {\em occurrence\/}
of $P$ at the position $i$ of the text $T$, if $P$ is a prefix of the
suffix $T[i,n]$, i.e., $P= T[i,i + p - 1]$.  A key observation is
that: {\em Searching for the occurrences of a pattern $P$ in $T$
amounts to retrieve all text suffixes that have the pattern $P$ as a
prefix}.  In this respect, the ordered set $\suf(T)$ exploits an
interesting property found by Manber and Myers~\cite{MM93}: {\em the
suffixes having prefix $P$ occupy a contiguous part of $\suf(T)$\/}.
In addition, the leftmost (resp.  rightmost) suffix of this contiguous
part {\em follows (resp. precedes) the lexicographic position of $P$
(resp. $P\#$) in the ordered set $\suf(T)$\/}.  To perform fast string
searches is then paramount to use a data structure that efficiently
retrieves the lexicographic position of a string in the ordered set
$\suf(T)$.

As an example, let us set $T=abababbc$ and consider the
lexicographically ordered set of all text suffixes $\suf(T)
=\{1,3,5,2,4,6,7,8\}$ (indicated by means of their starting positions
in $T$).  If we have $P=ab$, its lexicographic position in $\suf(T)$
precedes the first text suffix $T[1,8]=abababbc$, whereas the
lexicographic position of $P\#$ in $\suf(T)$ follows the fifth text
suffix $T[5,8] = abbc$.  From Manber-Myers' observation (above), the
three text suffixes between $T[1,8]$ and $T[5,8]$ in $\suf(T)$ are the
only ones prefixed by $P$ and thus $P$ occurs in $T$ three times at
positions $1, 3$ and $5$.  If we instead have $P=baa$, then both $P$
and $P\#$ have their lexicographic position in $\suf(T)$ between
$T[5,8]=abbc$ and $T[2,8]=bababbc$, so that $P$ does not occur in $T$.

The above definitions can be immediately extended to a text collection
$\Delta$ by replacing $\suf(T)$ with the set $\suf(\Delta)$ obtained
by merging lexicographically the suffixes in $\suf(S)$ for all texts
$S \in \Delta$.

\subsection{Suffix arrays and suffix trees}
\label{sa-st}

The suffix array~\cite{MM93}, or the PAT-array~\cite{pat-tree}, is an
indexing data structure that supports fast substring searches whose
cost does not depend on the alphabet's size.  A suffix array consists
of an array-based implementation of the set $\suf(T)$. In the example
above, the suffix array $SA$ equals to $[1,3,5,2,4,6,7,8]$. The search
in $T$ for an arbitrary pattern $P[1,p]$ exploits the lexicographic
order present in $SA$ and the two structural observations made
above. Indeed it first determines the lexicographic position of $P$ in
$\suf(T)$ via a binary search with {\em one level of indirection}: $P$
is compared against the text suffix pointed to by the examined $SA$'s
entry. Each pattern-suffix comparison needs $O(p)$ time in the worst
case, and thus $O(p \log n)$ time suffices for the overall binary
search. In our example, at the first step $P=ab$ is compared against
the entry $SA[4]=2$, i.e. the 2nd suffix of $T$, and the binary search
proceeds within the first half of $SA$ since $P \leq_L
T[2,8]=bababbc$. After that the lexicographic position of $P$ in $SA$
has been found, the search algorithm scans rightward the suffix array
until it encounters suffixes prefixed by $P$.  This takes $O(p \; occ)$
time in the worst case, where $occ$ is the number of occurrences of
$P$ in $T$. In our example, the lexicographic position of $P$ is
immediately before the first entry of $SA$, and there are three
suffixes prefixed by $P$ since $P$ is not a prefix of
$T[SA[4],8]=T[2,8]=bababbc$.

Of course the true behavior of the search algorithm depends on how
many long prefixes of $P$ occur in $T$. If there are very few of such
long prefixes, then it will rarely happen that a pattern-suffix
comparison in a binary-search step takes $\Theta(p)$ time, and
generally the $O(p \log n)$ bound is quite pessimistic. In ``random''
strings this algorithm requires $O(p + \log n)$ time. This latter
bound can be forced to hold in the worst case too, by adding an
auxiliary array, called $Lcp$ array, and designing a novel search
procedure~\cite{MM93}. The array $Lcp$ stores the
longest-common-prefix information between any two adjacent suffixes of
$\suf(T)$, thus it has the same length of $SA$. The novel search
procedure still proceeds via a binary search, but now a pattern-suffix
comparison does not start from the first character of the compared
strings but it takes advantage of the comparisons already executed and
the information available in the $Lcp$ array. However, since
practitioners prefer simplicity and space-compaction to
time-efficiency guarantee, this faster but space-consuming algorithm
is rarely used in practice. From a practical point of view, suffix
arrays are a much space-efficient full-text indexing data structure
because they store only one pointer per indexed suffix (i.e. usually
$3$ bytes suffice).  Nonetheless suffix arrays are pretty much static
and, in case of long text strings, the contiguous space needed for
storing them can become too constraining and may induce poor
performance in an external-memory setting. In fact, $SA$ can be easily
mapped onto disk by stuffing $\Theta(B)$ suffix pointers per
page~\cite{pat-tree}, but in this case the search bound is
$O(\frac{p}{B} \log_2 N + \frac{occ}{B})$ I/Os, and it is poor in
practice because all of these I/Os are {\em random}.

To remedy this situation~\cite{BYBZ96} proposed the use of {\em
supra-indices} over the suffix array. The key idea is to sample one
out of $b$ suffix array entries (usually $b = \Theta(B)$ and one entry
per disk page is sampled), and to store the first $\ell$ characters of
each sampled suffix in the supra-index. This supra-index is then used
as a first step to reduce the portion of the suffix array where the
binary search is performed. Such a reduction impacts favorably on the
overall number of random I/Os required by the search operation. Some
variations on this theme are possible, of course. For example the
supra-index does not need to sample the suffix array entries at fixed
intervals, and it does not need to copy in memory the same number
$\ell$ of suffix characters from each sampled suffix. Both these
quantities might be set according to the text structure and the space
available in internal memory for the supra-index. It goes without
saying that if the sampled suffixes are chosen to start at word
boundaries and entire words are copied into the supra-index, the
resulting data structure turns out to be {\em actually} an inverted
index. This shows the high flexibility of full-text indexing data
structures that, for a proper setting of their parameters, boil down
eventually to the weaker class of word-based indexes.

\begin{figure}
\centerline{\psfig{file=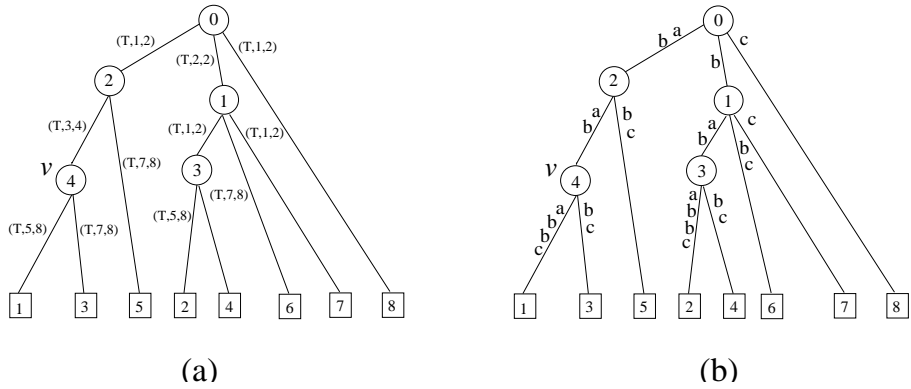,width=12cm}} 
\caption{\em (a)~The suffix tree for string $T=$~``abababbc''. We have
that node $v$ spells out the string `abab'. The substrings are
represented by triples to occupy constant space, each internal node
stores the length of its associated string, and each leaf stores the
starting position of its corresponding suffix. For our convenience, we
illustrate in~(b) the suffix tree showed in~(a) by explicitly writing
down the string $T[i,j]$ represented by the triple $(T,i,j)$. The
endmarker $\#$ is not shown. Reading the leaves rightward we get the
suffix array of $T$.\label{fig:suffix-tree}}
\end{figure}

On the other extreme, the smaller is the sampling step, the larger is
the memory requirement for the supra-index, and the faster is the
search operation. Sampling every suffix would be fabulous for query
performance but the quadratic space occupancy would make this approach
unaffordable. Actually if a compacted trie is used to store all the
suffixes, we end up into the most famous, elegant, powerful and widely
employed~\cite{Apo85,Gus97} full-text indexing data structure, known
as the {\em suffix tree}~\cite{McC76}.  Each arc of the suffix tree is
labeled with a text substring $T[i,j]$, represented via the triple
$(T,i,j)$, and the sibling arcs are ordered according to their first
characters, which are distinct (see Figure~\ref{fig:suffix-tree}).
There are no nodes having only one child except possibly the root and
each node has associated the string obtained by concatenating the
labels found along the downward path from the root to the node itself.
By appending the special character $\#$ to the text, the leaves have a
one-to-one correspondence to the text suffixes, each leaf stores a
different suffix and their rightward scanning gives actually the
suffix array. It is an interesting exercise to design an algorithm
which goes from the suffix array and the $Lcp$ array to the suffix
tree in linear time.

Suffix trees are also augmented by means of some special node-to-node
pointers, called {\em suffix links\/}~\cite{McC76}, which turn out to
be crucial for the efficiency of complex searches and updates. The
suffix link from a node storing a nonempty string, say $aS$ for a
character $a$, leads to the node storing $S$ and this node always
exists.  There can be $\Theta(|\Sigma|)$ suffix links leading to a
suffix-tree node because we can have one suffix link for each possible
character $a \in \Sigma$. Suffix trees require linear space and are
sometimes called {\em generalized\/} suffix trees when built upon a
text collection $\Delta$~\cite{AFILPS95,gst}. Suffix trees, and
compacted tries in general, are very efficient in searching an
arbitrary pattern string because the search is directed by the pattern
itself along a downward tree path starting from the root. This gives a
search time proportional to the pattern length, instead of a
logarithmic bound as it occurred for suffix arrays. Hence searching
for the $occ$ occurrences of a pattern $P[1,p]$ as a substring of
$\Delta$'s texts requires $O(p \log |\Sigma| + occ)$ time.  Inserting
a new text $T[1,m]$ into $\Delta$ or deleting an indexed text from
$\Delta$ takes $O(m \log |\Sigma|)$ time. The structure of a suffix
tree is rich of information so that statistics on text
substrings~\cite{Apo85} and numerous types of complex
queries~\cite{Gus97,navarro-survey} can be efficiently implemented.

Since the suffix tree is a powerful data structure, it would seem
appropriate to use it in external memory. To our surprise, however,
suffix trees loose their good searching and updating {\em worst-case}
performance when used for indexing large text collections that do not
fit into internal memory.  This is due to the following reasons:

\begin{itemize}
  
\item[a.] Suffix trees have an {\em unbalanced topology\/} that is
text-dependent because their internal nodes are in correspondence to
some repeated substrings.  Consequently, these trees inevitably
inherit the drawbacks pointed out in scientific literature with regard
to paging unbalanced trees in external memory.  There are some good
average-case solutions to this problem that group $\Theta(B)$ nodes
per page under node insertions only~\cite[Sect.6.2.4]{Knuth:1998:SS}
(deletions make the analysis extremely difficult~\cite{Sprugnoli}),
but they cannot avoid storing a downward path of $k$ nodes in
$\Omega(k)$ {\em distinct\/} pages in the worst case.
  
\item[b.] Since the outdegree of a node can be $\Theta(|\Sigma|)$, its
pointers to children might not fit into $O(1)$ disk pages so they
would have to be stored in a separate B-tree.  This causes an
$O(\log_B |\Sigma|)$ disk access overhead for each branch out of a
node both in searching and updating operations.
  
\item[c.] Branching from a node to one of its children requires
further disk accesses in order to retrieve the disk pages containing
the substring that labels the traversed arc.
  
\item[d.] Updating suffix trees under string insertions or
deletions~\cite{AFILPS95,gst} requires the insertion or deletion of
some nodes in their unbalanced structure. This operation inevitably
relies on merging and splitting disk pages in order to occupy
$\Theta(\frac{N}{B})$ of them.  This approach is very expensive:
splitting or merging a disk page can take $O(B |\Sigma|)$ disk
accesses because $\Theta(B)$ nodes can move from one page to another.
The $\Theta(|\Sigma|)$ suffix links leading to each moved node must be
redirected and they can be contained in different pages.

\end{itemize}

Hence we can conclude that, if the text collection $\Delta$ is stored
on disk, the search for a pattern $P[1,p]$ as a substring of
$\Delta$'s texts takes $O(p \log_B |\Sigma| + occ)$ worst-case disk
accesses (according to Points~a--c).  Inserting an $m$-length text in
$\Delta$ or deleting an $m$-length text from $\Delta$ takes $O(m B
|\Sigma|)$ disk accesses in the worst-case (there can be $\Theta(m)$
page splits or merges, according to point~(d)).

From the point of view of average-case analysis, suffix tree and
compacted trie performances in external memory are heuristic and
usually confirmed by
experimentation~\cite{AoeMSP96,MerSha93,Munro98,DeJTanVan87,lc-trie}. The
best result to date is the {\em Compact PAT-tree}~\cite{CM96}. It is a
succinct representation of the (binary) Patricia
tree~\cite{Morrison68}, it occupies about $5$ bytes per suffix and
requires about $5$ disk accesses to search for a pattern in a text
collection of $100$Mb. The paging strategy proposed to store the
Compact PAT-tree on disk is a heuristic that achieves only $40\%$ page
occupancy and slow update performance~\cite{CM96}. From the
theoretical point of view, pattern searches require
$O(\frac{h}{\sqrt{p}} + \log_p N)$ I/Os, where $h$ is the Patricia
tree's height; inserting or deleting a text in $\Delta$ costs at least
as searching {\em for all} of its suffixes individually.  Therefore
this solution is attractive only in practice and for static textual
archives. Another interesting implementation of suffix trees has been
proposed in~\cite{Kurtz-st}. Here the space occupancy has been
confined between $10$ and $20$ bytes per text suffix, assuming a text
shorter than $2^{27}$ characters.

\subsection{Hybrid data structures}
\label{sub:Hybr-data-struct}

Although suffix arrays and compacted tries present good properties,
none of them is explicitly designed to work on a hierarchy of memory
levels. The simple paging heuristics shown above are not acceptable
when dealing with large text collections which extensively and
randomly access the external storage devices for both searching or
updating operations.  This is the reason why various researchers have
tried to properly combine these two approaches in the light of the
characteristics of the current hierarchy of memory levels.  The result
is a family of {\em hybrid data structures} which can be divided into
two large subclasses.

One subclass contains data structures that exploit the no longer
negligible size of the internal memory of current computers by keeping
{\em two indexing levels}: one level consists of a compacted trie (or
a variant of it) built on a {\em subset} of the text suffixes and
stored in internal memory (previously called {\em supra-index}); the
other level is just a plain suffix array built over all the suffixes
of the indexed text.  The trie is used to route the search on a small
portion of the suffix array, by exploiting the efficient random-access
time of internal memory; an external-memory binary search is
subsequently performed on a restricted part of the suffix array, so
identified, thus requiring a reduced number of disk accesses.  Various
approaches to suffix sampling have been introduced in the
literature~\cite{ColussiC1996,Kaerkkaeinen:1996:SST,Munro98,CPM::AnderssonLS1996},
as well various trie coding methods have been employed to stuff as
much suffixes as possible into internal
memory~\cite{BYBZ96,lc-trie,DeJTanVan87,CPM::Karkkainen1995}.  In all
these cases the aim has been to balance the efficient search
performance of compacted tries with the small space occupancy of
suffix arrays, taking into account the limited space available into
internal memory.  The result is that: (1)~the search time is faster
than in suffix arrays (see e.g.~\cite{BYBZ96,CPM::AnderssonLS1996})
but it is yet not optimal because of the binary search on disk,
(2)~the updates are slow because of the external-memory suffix array,
and (3)~slightly more space is needed because of the internal-memory
trie.

The second subclass of hybrid data structures has been obtained by
properly combining the B-tree data structure~\cite{Comer:1979:UBT}
with the effective routing properties of suffix arrays, tries or their
variants. An example is the Prefix B-tree~\cite{Bayer:1977:PB} that
explicitly stores prefixes of the indexed suffixes (or indexed
strings) as routing information (they are called {\em separators})
into its internal nodes. This design choice poses some algorithmic
constraints. In fact the updates of Prefix B-trees are complex because
of the presence of arbitrarily long separators, which require
recalculations and possibly trigger new expansions/contractions of the
B-tree nodes.  Various works have investigated the splitting of Prefix
B-tree nodes when dealing with variable length
keys~\cite{Bayer:1977:PB,lncs303*309} but all of them have been faced
with the problem of choosing a proper splitting separator.  For these
reasons, while B-trees and their basic variants are among the most
used data structures for primary key
retrieval~\cite{Comer:1979:UBT,Knuth:1998:SS}, Prefix B-trees are not
a common choice as full-text indices because their performance is
known to be not efficient enough when dealing with arbitrarily long
keys or highly dynamic environments.

\subsection{The string B-tree data structure}
\label{sb-tree}

The String B-tree~\cite{FG:99} is a hybrid data structure introduced
to overcome the limitations and drawbacks of Prefix B-trees. The key
idea is to plug a Patricia tree~\cite{Morrison68} into the nodes of
the B-tree, thus providing a {\em routing tool} that efficiently
drives the subsequent searches and, more importantly, occupies a space
proportional to the number of indexed strings instead of their total
length.  The String B-tree achieves optimal search bounds (in the case
of an unbounded alphabet) and attractive update performance. In
practice it requires a negligible, guaranteed, number of disk accesses
to search for an arbitrary pattern string in a large text collection,
independent of the character distribution.  We now recall the main
ideas underlying the String B-tree data structure. For more
theoretical details we refer the reader to~\cite{FG:99}, for a
practical analysis we refer to~\cite{FG:96} and
Section~\ref{engineering}.

String B-trees are similar to B$^+$-trees~\cite{Comer:1979:UBT}, the
keys are {\em pointers\/} to the strings in $\suf(\Delta)$ (i.e. to
suffixes of $\Delta$'s strings), they reside in the leaves and some
copies of these keys are stored in the internal nodes for routing the
subsequent traversals.  The order between any two keys is the
lexicographic order among the corresponding pointed strings.  The
novelty of the String B-tree is that the keys in each node are not
explicitly stored, so that they may be of arbitrary length.  Only the
string pointers are kept into the nodes, organized by means of a
Patricia tree~\cite{Morrison68} which ensures small overhead in
routing string searches or updates, and occupies space proportional to
the {\em number} of indexed strings rather than to their total length.

\begin{figure}[t]
\centerline{\psfig{file=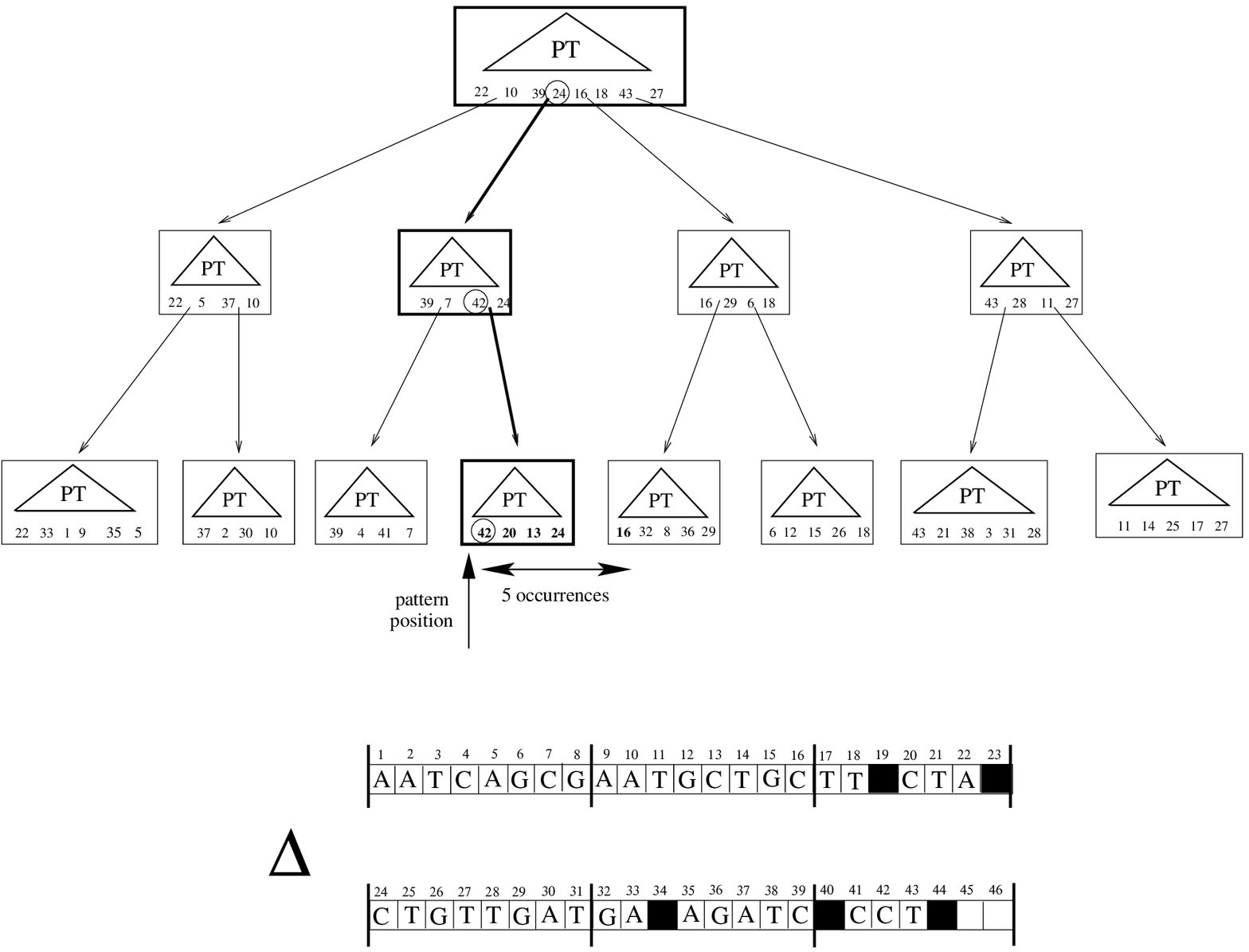,width=13.5cm}} 
\caption{\em An illustrative example depicting a String B-tree built
on a set $\Delta$ of DNA sequences. $\Delta$'s strings are stored in a
file separated by special characters, here denoted with black
boxes. The triangles labeled with $PT$ depict the Patricia trees
stored into each String B-tree node.  The figure also shows in bold
the String B-tree nodes traversed by the search for a pattern $P=
``CT''$. The circled pointers denote the suffixes, one per level,
explicitly checked during the search; the pointers in bold, in the
leaf level, denote the five suffixes prefixed by $P$ and thus the five
positions where $P$ occurs in $\Delta$. \label{fig:sb-tree}}
\end{figure}

We denote by $SBT_\Delta$ the string B-tree built on the text
collection $\Delta$, and we adopt two conventions: there is {\em no
distinction\/} between a key and its corresponding pointed string;
each disk page can contain up to $2b$ keys, where $b=\Theta(B)$ is a
parameter depending on the actual space occupancy of a node (this will
be discussed in Section~\ref{engineering}).  In detail, the strings of
$\suf(\Delta)$ are distributed among the String B-tree nodes as shown
in Figure~\ref{fig:sb-tree}.  $\suf(\Delta)$ is partitioned into
groups of at most $2b$ strings each (except the last group which may
contain fewer strings) and every group is stored into a leaf of
$SBT_\Delta$ in such a way that the left-to-right scanning of these
leaves gives the ordered set $\suf(\Delta)$ (i.e. the suffix array of
$\Delta$).  Each internal node $\pi$ has $n(\pi)$ children, with
$\frac{b}{2} \leq n(\pi) \leq b$ (except the root which has from $2$
to $b$ children). Node $\pi$ also stores the string set $\S_\pi$
formed by copying the leftmost and the rightmost strings contained in
each of its children. As a result, set $\S_\pi$ consists of $2n(\pi)$
strings, node $\pi$ has $n(\pi)=\Theta(B)$ children, and thus the
height of $SBT_\Delta$ is $O(\log_B N)$ where $N$ is the total length
of $\Delta$'s strings, or equivalently, the cardinality of
$\suf(\Delta)$.

The main advantage of String B-trees is that they support the standard
B-tree operations, now, on arbitrary long keys. Since the String
B-tree leaves form a suffix array on $\suf(\Delta)$, the search for a
pattern string $P[1,p]$ in $SBT_\Delta$ must identify foremost the
lexicographic position of $P$ among the text suffixes in
$\suf(\Delta)$, and thus, among the text pointers in the String B-tree
leaves. Once this position is known, all the occurrences of $P$ as a
substring of $\Delta$'s strings are given by the consecutive pointers
to text suffixes which start from that position and have $P$ as a
prefix (refer to the observation on suffix arrays, in
Section~\ref{full-text}). Their retrieval takes $O((p/B) occ)$ I/Os,
in case of a brute-force match between the pattern $P$ and the checked
suffixes; or the optimal $O(occ/B)$ I/Os, if some additional
information about the longest-common-prefix length shared by adjacent
suffixes is kept into each String B-tree leaf.  In the example of
Figure~\ref{fig:sb-tree} the search for the pattern $P = ``CT''$
traces a downward path of String B-tree nodes and identifies the
lexicographic position of $P$ into the fourth String B-tree leaf (from
the left) and before the $42$th text suffix. The pattern occurrences
are then retrieved by scanning the String B-tree leaves from that
position until the $32$th text suffix is encountered, because it is
not prefixed by $P$. The text positions $\{42,20,13,24,16\}$ denote
the five occurrences of $P$ as a substring of $\Delta$'s texts.

\begin{figure}[t]
\centerline{\psfig{file=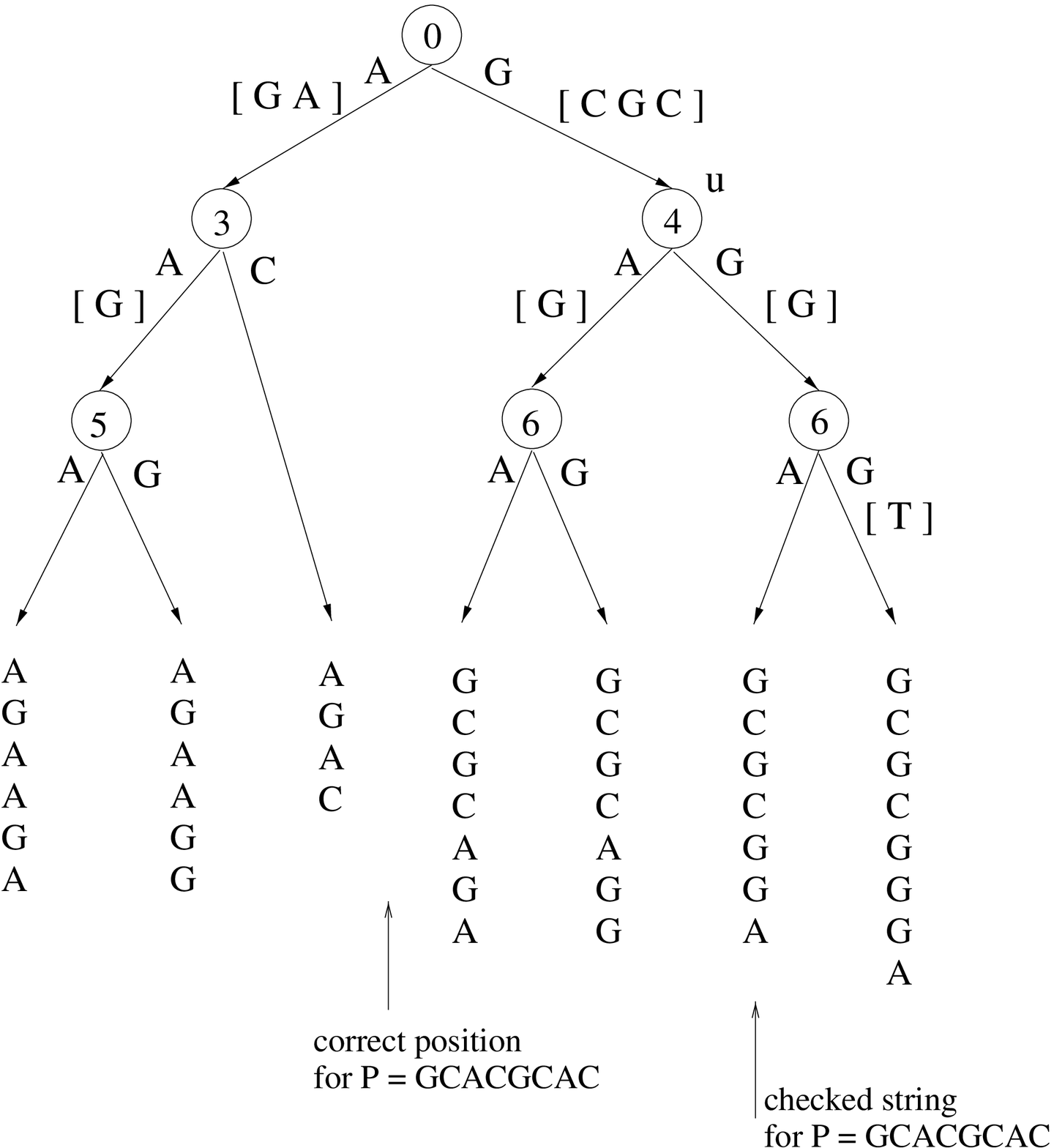,width=8cm}} 
\caption{\em An example of Patricia tree built on a set of $k=7$ DNA
strings drawn from the alphabet $\Sigma = \{A, G, C, T\}$.  Each leaf
points to one of the $k$ strings; each internal node $u$ (they are at
most $k-1$) is labeled with one integer $len(u)$ which denotes the
length of the common prefix shared by all the strings pointed by the
leaves descending from $u$; each arc (they are at most $2k-1$) is
labeled with only one character (called {\em branching
character}). The characters between square-brackets are not explicitly
stored, and denote the other characters labeling a trie
arc. \label{fig:patrie}}
\end{figure}

Therefore the efficient implementation of string searches in String
B-trees boils down to the efficient routing of the pattern search
among the String B-tree nodes. In this respect it is clear that the
way a string set $\S_\pi$, in each traversed node $\pi$, is organized
plays a crucial role.  The innovative idea in String B-trees is to use
a Patricia tree $PT_\pi$ to organize the string pointers in
$\S_\pi$~\cite{Morrison68}.  Patricia trees preserve the searching
power and properties of compacted tries, although in a reduced space
occupancy. In fact $PT_\pi$ is a simplified trie in which each arc
label is replaced by only its first character. See
Figure~\ref{fig:patrie} for an illustrative example.

When the String B-tree is traversed downward starting from the root,
the traversal is routed by using the Patricia tree $PT_\pi$ stored in
each visited node $\pi$. The goal of $PT_\pi$ is to help finding the
lexicographic position of the searched pattern $P$ in the ordered set
$\S_\pi$. This search is a little bit more complicated than the one in
classical tries (and suffix trees), because of the presence of only
one character per arc label, and in fact consists of two stages:

\begin{itemize}
  
\item Trace a downward path in $PT_{\pi}$ to locate a leaf $l$ which
points to an {\em interesting} string of $\S_\pi$. This string does
not necessarily identify $P$'s position in $\S_\pi$ (which is our
goal), but it provides enough information to find that position in the
second stage (see Figure~\ref{fig:patrie}). The retrieval of the
interesting leaf $l$ is done by traversing $PT_\pi$ from the root and
comparing the characters of $P$ with the single characters which label
the traversed arcs until a leaf is reached or no further branching is
possible (in this case, choose $l$ to be any descendant leaf from the
last traversed node).
  
\item Compare the string pointed by $l$ with $P$ in order to determine
their longest common prefix.  A useful property holds~\cite{FG:99}:
{\em the leaf $l$ stores one of the strings in $\S_\pi$ that share the
{\bf longest\/} common prefix with $P$.\/} The length $\ell$ of this
common prefix and the mismatch character $P[\ell+1]$ are used in two
ways: first to determine the shallowest ancestor of $l$ spelling out a
string longer than $\ell$; and then, to select the leaf descending
from that ancestor which identifies the lexicographic position of $P$
in $\S_\pi$.

\end{itemize}

An illustrative example of a search in a Patricia tree is shown in
Figure~\ref{fig:patrie} for a pattern $P = ``GCACGCAC''$. The leaf $l$
found after the first stage is the second one from the right. In the
second stage, the algorithm first computes $\ell = 2$ and $P[\ell+1] =
A$; then, it proceeds along the leftmost path descending from the node
$u$, since the 3rd character on the arc leading to $u$ (i.e. the
mismatch $G$) is grater than the corresponding pattern character $A$.
The position reached by this two-stage process is indicated in
Figure~\ref{fig:patrie}, and results the correct lexicographic
position of $P$ among $\S_\pi$'s strings.

We remark here that $PT_\pi$ requires space linear in the {\em
number\/} of strings of $\S_\pi$, therefore the space usage is
independent of their {\em total length\/}. Consequently, the number of
strings in $\S_\pi$ can be properly chosen in order to be able to fit
$PT_\pi$ in the disk page allocated for $\pi$. An additional nice
property of $PT_\pi$ is that it allows to find the lexicographic
position of $P$ in $\S_\pi$ by exploiting the information
available in $\pi$'s page and by fully comparing $P$ with just {\em
one of the strings in $\S_\pi$\/}. This clearly allows to reduce the
number of disk accesses needed in the routing step. By counting the
number of disk accesses required for searching $P[1,p]$ in the strings
of $\Delta$, and recalling that $\Delta$'s strings have overall length
$N$, we get the I/O-bound $O(\frac{p}{B} \log_B N)$. In fact,
$SBT_\Delta$ has height $O(\log_B N)$, and at each traversed node
$\pi$ we may need to fully compare $P$ against one string of $\S_\pi$
thus taking $O(\frac{p}{B} + 1)$ disk accesses.

A further refinement to this idea is possible, thought, by observing
that we do not necessarily need to compare the two strings, i.e. $P$
and the candidate string of $\S_\pi$, starting from their first
character but we can take advantage of the comparisons executed on the
ancestors of $\pi$, thus skipping some character comparisons and
reducing the number of disk accesses.  An incremental accounting
strategy allows to prove that $O(\frac{p}{B} + \log_B N)$ disk
accesses are indeed sufficient, and this bound is optimal in the case
of an unbounded alphabet. A more complete analysis and description of
the search and update operations is given in~\cite{FG:99} where it is
formally proved the following:

\begin{theorem}
\label{teo:jacm}
String B-trees support the search for all the $occ$ occurrences of an
arbitrary pattern $P[1,p]$ in the strings of a set $\Delta$ taking
$O(\frac{p+occ}{B}+\log_B N)$ disk accesses, where $N$ is the overall
length of $\Delta$'s strings. The insertion or the deletion of an
$m$-length string in/from the set $\Delta$ takes $O(m \log_B (N+m))$
disk accesses.  The required space is $\Theta(\frac{N}{B})$ disk
pages.
\end{theorem}

As a corollary, we get a result which points out the String B-tree as
an effective data structure also for dictionary applications.

\begin{corollary}
\label{teo:jacm2}
String B-trees support the search for all the $occ$ occurrences of an
arbitrary pattern $P[1,p]$ as a prefix of the $K$ strings in a set
$\Delta$ taking $O(\frac{p+occ}{B}+\log_B K)$ disk accesses. The
insertion or the deletion of an $m$-length string in/from the set
$\Delta$ takes $O(\frac{m}{B} + \log_B K)$ disk accesses.  The space
usage of the String B-tree is $\Theta(\frac{K}{B})$ disk pages,
whereas the space occupied by the string set $\Delta$ is
$\Theta(\frac{N}{B})$ disk pages.
\end{corollary}

Some authors have successfully used String B-trees in other settings:
multi-dimensional prefix-string queries~\cite{divesh-nick},
conjunctive boolean queries on two substrings~\cite{pods01},
dictionary matching problems~\cite{IC98}, distributed search
engines~\cite{ferra-luccio}, indexing of XML texts~\cite{CFP99}. All
of these applications show the flexibility of this data structure, its
efficiency in external memory, and foretell engineered implementations
because up to now String B-trees have been confined mainly to the
theoretical realm perhaps because of their space occupancy: the best
known implementation uses about $12$ bytes per indexed
suffix~\cite{FG:96}.  Given this bottleneck, less I/O-efficient but
space cheaper data structures have been preferred in practice
(e.g. supra-indexes~\cite{BYBZ96}). In the next section we try to
overcome this limitation by proposing a novel engineered version of
String B-trees suitable for practical implementations.

\subsection{Engineering the String B-tree}
\label{engineering}

String B-trees have the characteristics that their height decreases
exponentially as $b$'s value increases (with fixed $N$).  The value of
$b$ is strictly related to the number of strings contained in each
node $\pi$ because $b \leq |\S_\pi| \leq 2b$.  If the disk page size
$B$ increases, we can store more suffixes in $\S_\pi$.  However, since
$B$ is typically chosen to be proportional to the size of a disk page,
we need a technique that maximizes $|\S_\pi|$ for a fixed disk page
size $B$.

The space occupancy of a String B-tree node $\pi$ is evaluated as the
sum of three quantities:

\begin{enumerate}

\item The amount of auxiliary and bookkeeping information necessary to
node $\pi$. This is practically negligible and, hereafter, it will not
be accounted for.
  
\item The amount of space needed to store the pointers to the children
of $\pi$. This quantity is absent for the leaves; in the case of
internal nodes, usually a $4$-byte pointer suffices.
  
\item The amount of space required to store the pointers to the
strings in $\S_{\pi}$ and the associated machinery $PT_\pi$.  This
space is highly implementation dependent, so deserves an accurate
discussion.

\end{enumerate}

Let us therefore concentrate on the amount of space required to store
$\S_{\pi}$ and $PT_{\pi}$. This is determined by three kinds of
information: (i) the Patricia tree topology, (ii) the integer values
kept into the internal nodes of $PT_{\pi}$ (denoted by $len$), and
(iii) the pointers to the strings in $\S_{\pi}$.  The na\"{\i}ve
approach to implement (i--iii) is to use {\em explicit pointers} to
represent the parent-child relationships in $PT_\pi$ and the strings
in $\S_\pi$, and allocate $4$ bytes for the $len$ values.  Although
simple and efficient in supporting search and update operations, this
implementation induces an unacceptable space occupancy of about $24$
bytes per string of $\S_\pi$~!  The literature about space-efficient
implementations of Patricia trees is huge but some ``pruning'' of
known results can be done according to the features of our trie
encoding problem. {\em Hash-based} representation of
tries~\cite{Darragh:1993:BCR}, although elegant and succinct, can be
discarded because they do not have guaranteed performance in time and
space, and they are not better than classical tries on small string
sets~\cite{cached-trie,bs}, as it occurs in our $\S_\pi$'s sets. {\em
List or array-based} implementations of Patricia trees adopting path
and/or level compression
strategies~\cite{lc-trie,Andersson:1993:IBT,LCP-trie98} are space
consuming and effective mainly on random data.

More appealing for our purposes is a recent line of research pioneered
by~\cite{FOCS::Jacobson1989} and extended by other
authors~\cite{Munro97,Munro98,CM96,KatMak90a,Makinen:1991:SBT} to the
succinct encoding of Patricia trees. Their main idea is to succinctly
encode the Patricia tree topology and then use some other data
structures to properly encode the other information, like the string
pointers (kept into the leaves) and the $len$ values (kept into the
internal nodes). The general policy is therefore to handle the data
and the tree structure separately.  This enables to compress the plain
data using any of the known methods (see e.g.~\cite{wmb99}) and
independently find an efficient coding method for the tree structure
irrespective of the form and contents of the data items stored in its
nodes and leaves.

In the original implementation of String B-trees~\cite{FG:96}, the
shape of $PT_\pi$ was succinctly encoded via two operations, called
{\em compress} and {\em uncompress}. These operations allow to go from
a Patricia tree to a binary sequence, and vice versa, by means of a
preorder traversal of $PT_\pi$. Although space efficient and simple,
this encoding is CPU-intensive to be updated or searched, so that a
small page size of $B = 1$ kilobytes was chosen in~\cite{FG:96} to
balance the CPU-cost of node compression/uncompression and the
I/O-cost of the update operations (see~\cite{FG:96} for details). Here
we propose a novel encoding scheme that surprisingly throws away the
Patricia tree topology, keeps just the string pointers and the $len$
values, and is still able to support pattern searches in a constant
number of I/Os per visited String B-tree node. As a result, the
asymptotic I/O-bounds stated in Theorem~\ref{teo:jacm} still hold with
a significant space improvement in the constants hidden in the big-Oh
notation.

The starting point is the beautiful result of~\cite{Ferguson:1992:BTD}
that we briefly recall here. Let us be given a lexicographically
ordered array of string pointers, called $SP$, and the array of
longest-common-prefixes shared by strings adjacent in $SP$, called
$Lcp$. We can look at $SP$ and $Lcp$ as the sequence of string
pointers and $len$ values encountered in an inorder traversal of the
Patricia tree $PT_\pi$ stored into a given String B-tree node
$\pi$. Now, let us assume that we wish to route the search for a
pattern $P[1,p]$ through node $\pi$, we then need to find the
lexicographic position of $P$ in $SP$ since it indexes $\S_\pi$. We
might implement that search via the classical binary search procedure
on suffix arrays within a logarithmic number of I/Os (see
Section~\ref{sa-st}).  The result in~\cite{Ferguson:1992:BTD} shows
instead that it is enough to execute only {\em one} string access, few
more $\Theta(p + k)$ bit comparisons and one full scan of the arrays
$Lcp$ and $SP$. Of course this new algorithm is unaffordable on large
arrays, but this is not our context of application: the string set
$\S_\pi$ actually consists of {\em few thousands} of items (stored in
one disk page), and the arrays $SP$ and $Lcp$ reside in memory when
the search is performed (i.e. the disk page has been fetched). Hence
the search is I/O-cheap in that it requires just one sequential string
access, it is CPU-effective because the array-scan can benefit from
the reading-ahead policy of the internal cache, and is space efficient
because it avoids the storage of $PT_\pi$'s topology.

\begin{figure}[t]
\centerline{\psfig{file=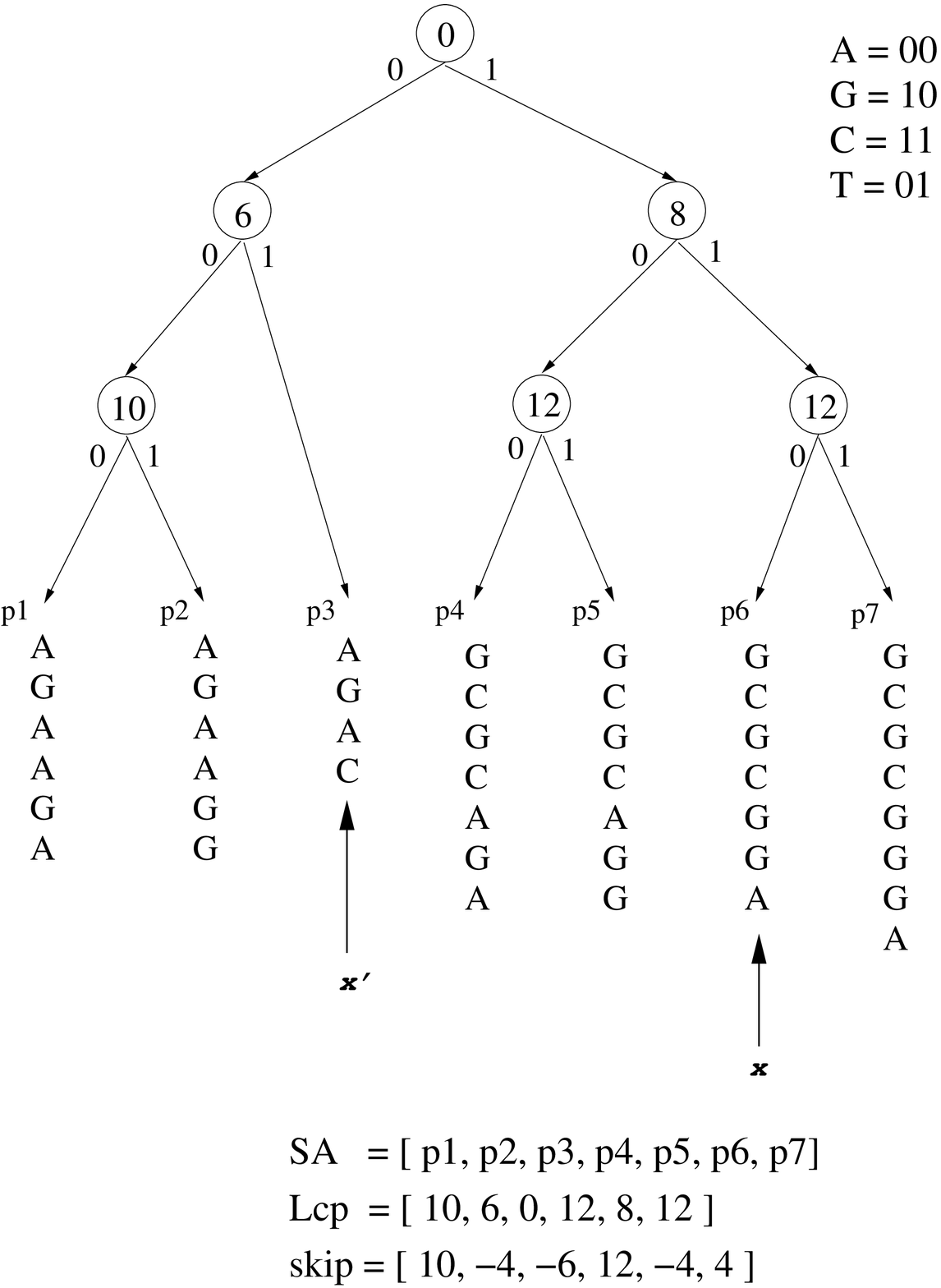,width=8cm}} 
\caption{\em The arrays $SP$ and $Lcp$ computed on the Patricia tree
of Figure~\protect\ref{fig:patrie}. The array $Skip$ is derived from
the array $Lcp$ by subtracting its adjacent entries. The $Skip$s and
$Lcp$s are expressed in bits.\label{fig:ferguson}}
\end{figure}

Let us therefore detail the search algorithm which assumes a {\em
binary} pattern $P$ and consists of two phases
(see~\cite{Ferguson:1992:BTD} for the uneasy proof of correctness). In
the first phase, the algorithm scans rightward the array $SP$ and
inductively keeps $x$ as the position of $P$ in this array (initially
$x=0$).  At a generic step $i$ it computes $\ell=Lcp[i]$, as the
mismatching position between the two adjacent strings $SP[i]$ and
$SP[i+1]$. Notice that the $\ell$th bit of the string $SP[i]$ is
surely 0, whereas the $\ell$th bit of the string $SP[i+1]$ is surely 1
because they are binary and lexicographically ordered. Hence the
algorithm sets $x=i+1$ and increments $i$ if $P[\ell]=1$; otherwise
(i.e. $P[\ell]=0$), it leaves $x$ unchanged and increments $i$ until
it meets an index $i$ such that $Lcp[i] < \ell$.  Actually, in this
latter case the algorithm is jumping all the succeeding strings which
have the $\ell$th bit set to $1$ (since $P[\ell]=0$).  The first phase
ends when $i$ reaches the end of $SP$; it is possible to prove that
$SP[x]$ is one of the strings in $SP$ sharing the longest common
prefix with $P$. In the illustrative example of
Figure~\ref{fig:ferguson}, we have $P = ``GCACGCAC''$ and coded its
characters in binary; the first phase ends by computing $x=4$. The
second phase of the search algorithm initiates by computing the length
$\ell'$ of the longest common prefix between $P$ and the candidate
string $SP[x]$. If $SP[x]=P$ then it stops, otherwise the algorithm
starts from position $x$ a backward scanning of $SP$ if $P[\ell'+1]=0$
or a forward scanning if $P[\ell'+1]=1$. This scan searches for the
lexicographic position of $P$ in $SP$ and proceeds until is met the
position $x'$ such that $Lcp[x']< \ell'$. The searched position lies
between the two strings $SP[x']$ and $SP[x'+1]$. In the example of
Figure~\ref{fig:ferguson}, it is $\ell'=4$ (in bits) and $P[5]=0$ (the
first bit of A's binary code); hence $SP$ is scanned backward from
$SP[4]$ for just one step since $Lcp[3]= 0 < 4=\ell'$. This is the
correct position of $P$ among the strings indexed by $SP$.

Notice that the algorithm needs to access the disk just for fetching
the string $SP[x]$ and comparing it against $P$. Hence $O(p/B)$ I/Os
suffice to route $P$ through the String B-tree node $\pi$. An
incremental accounting strategy, as the one devised in~\cite{FG:99},
allows to prove that we can skip some character comparisons and
therefore require $O(\frac{p+occ}{B} + \log_B N)$ I/Os to search for
the $occ$ occurrences of a pattern $P[1,p]$ as a substring of
$\Delta$'s strings. Preliminary experiments have shown that searching
few thousands of strings via this approach needs about $200\mu\!s$,
which is negligible compared to the $5.000\mu\!s$ required by a single
I/O on modern disks. Furthermore, the incremental search allows
sometimes to avoid the I/Os needed to access $SP[x]$~!

Some improvements to this idea are still possible both in time and
space. First, we can reduce the CPU-time of search and update
operations by adopting a sort of supra-index on $SP$ defined as
follows. We decompose the array $SP$ (and hence $Lcp$) into sub-arrays
of size $\Theta(\log^2 |SP|)$. The rightmost string of each subarray
is stored in a pointer-based Patricia tree. This way, the (sampled)
Patricia tree is used to determine the subarray containing the
position of the searched pattern; then the search procedure above is
applied to that subarray to find the correct position of $P$ into
it. The overall time complexity is $O(p)$ to traverse the Patricia
tree, and $O(p+\log^2 |SP|)$ to explore the reached sub-array. Notice
also that only two strings in $SP$ are accessed on disk. The data
structure is dynamic and every insertion or deletion of an $m$-length
string takes $O(m+\log^2 |SP|)$ time and only two string accesses to
the disk. The resulting data structure turns out to be simple, its
construction from scratch is fast and thus split/merge operations on
String B-tree nodes should be effective if $PT_\pi$ is implemented in
this way.

We point out that due to the sequential access to the array $Lcp$, a
further space saving is possible. We can compactly encode the entries
of array $Lcp$ by representing only their {\em differences}.  Namely,
we use a novel array $Skip$ in which each value denotes the difference
between two consecutive $Lcp$'s entries (i.e. $Skip[i] = Lcp[i] -
Lcp[i-1]$, see Figure~\ref{fig:ferguson}). Various experimental
studies on the distribution of the {\em Skip\/}s over standard text
collections have shown that most of them (about $90\%$~\cite{shang})
are small and thus they are suitably represented via variable-length
codes~\cite{CM96,MerSha93}. We suggest the use of the {\em
continuation bit} code, described in Section~\ref{word-based}, because
of two facts: the string sampling at the internal nodes of
$SBT_\Delta$ and the results in~\cite{shang} drives us to conjecture
small skips and thus one byte coding for them; furthermore, this
coding scheme is simple to be programmed, induces byte-aligned codes
and hence it is CPU efficient.

\medskip
We conclude this section by observing that up to now we assumed the
text collection $\Delta$ to be fixed.  In a real-life context, we
should expect that new texts are added to the collection and old texts
are removed from it. While handling deletions is not really a problem
as we have a plethora of tools inherited from standard B-trees,
implementing the addition of a new text requires decisely new
techniques.  This asymmetry between deletion and insertion is better
understood if we observe that the insertion of a new text $T[1,m]$
into $\Delta$ requires the insertion of all of its $m$ suffixes $\{
T[1,m], T[2,m], \ldots, T[m,m] \}$ into the lexicographically ordered
set $\suf(\Delta)$.  Consequently, the dominant cost is due to the
comparison of all characters in each text suffix that may sum up to
$\Theta(m^2)$. Since $T$ can be as large as $m=10^6$ characters (or
even more), the {\em rescanning} of the text characters might be a
computational bottleneck. On the other hand, the deletion of a text
$T[1,m]$ from $\Delta$ consists of a sequence of $m$ standard
deletions of $T$'s suffix pointers, and hence can exploit standard
B-tree techniques.

The approach proposed in~\cite{FG:99} to avoid the ``rescanning'' in
text insertion is mainly theoretical in its flavor and considers an
{\em augmented} String B-tree where some pointers are added to its
leaves.  The counterpart for this I/O improvement is that a larger
space occupancy is needed and, when rebalancing the String B-tree, the
redirection of some of these additional pointers may cause the
execution of random I/Os. Therefore, it is questionable if this
approach is really attractive from a practical point of view.
Starting from these considerations~\cite{FG:96} proposed an
alternative approach based on a {\em batched\/} insertion of the $m$
suffixes of $T$. This approach exploits the LRU~buffering strategy of
the underlying operating system and proves effective in the case of a
large $m$. In the case of a small $m$ a different approach must be
adopted which is based on the {\em suffix-array merging} procedure
presented in~\cite{pat-tree}: a suffix array $SA$ is built for $T$,
together with its $Lcp$ array; the suffix array $SA_\Delta$ on the
suffixes in $\suf(\Delta)$ is instead derived from the leaves of
$SBT_\Delta$ within $O(N/B)$ I/Os. The merge of $SA$ and $SA_\Delta$
(and their corresponding $Lcp$ arrays) gives the new set of String
B-tree leaves, the internal nodes are constructed within $O(N/B)$ I/Os
via the simple approach devised in Section~\ref{sb-tree}. Even if the
merging of the two suffix arrays can be dramatically slow in theory,
since every suffix comparison might require one disk access, the
character distribution of real text collections makes the $Lcp$ arrays
very helpful and allows to solve in practice most of the suffix
comparisons without accessing the disk. A throughtful sperimentation of
these approaches is still needed to validate such empirical
considerations.

\subsection{String B-tree construction}
\label{full:construction}

The efficient construction of full-text indexes on very large text
collections is a hot topic: ``{\em We have seen many papers in which
the index simply `is', without discussion of how it was created. But
for an indexing scheme to be useful it must be possible for the index
to be constructed in a reasonable amount of time,
.....}''~\cite{guidelines}. The construction phase may be, in fact, a
bottleneck that can prevent these powerful indexing tools to be used
even in medium-scale applications. Known construction algorithms are
very fast when employed on textual data that fit in the internal
memory of computers~\cite{MM93,S98:COSTR,Kurtz-st,FM:02SA} but their
performance immediately degrades when the text size becomes so large
that the texts must be arranged on (slow) external storage devices. In
the previous section we have addressed the problem of updating the
String B-tree under the insertion/deletion of a single text. Obviously
those algorithms cannot be adopted to construct from scratch the
String B-tree over a largely populated text collection because they
would incur in an enormous amount of random I/Os. In this section we
describe first an efficient algorithm to build the suffix array
$SA_\Delta$ for a text collection $\Delta$ of size $N$, and then
present a simple algorithm which derives the String B-tree
$SBT_\Delta$ from this array in $O(N/B)$ I/Os. For further theoretical
and experimental results on this interesting topic we refer the reader
to~\cite{FFM:00,CF99,S98:COSTR,pat-tree}.

\medskip
{\bf How to build $SA_\Delta$. } As shown in~\cite{CF99}, the most
attractive algorithm for building large suffix arrays is the one
proposed in~\cite{pat-tree} because it requires only $4$ bytes of
working space per indexed suffix, it accesses the disk mostly in a
sequential manner and it is very simple to be programmed.  For the
simplicity of presentation, let us assume to concatenate all the texts
in $\Delta$ into just one single long text $T$ of length $N$, and let
us concentrate on the construction of the suffix array $SA_T$ of
$T$. The transformation from $SA_T$ to $SA_\Delta$ is easy and left to
the reader as an exercise.

The algorithm computes incrementally the suffix array $SA_T$ in
$\Theta(N/M)$ stages. Let $\ell < 1$ be a positive constant fixed
below, and assume to set a parameter $m= \ell M$ which, for the sake
of presentation, divides $N$.  This parameter will denote the size of
the text pieces loaded in memory at each stage. 

The algorithm maintains at each stage the following invariant: {\em At
the beginning of stage $h$, with $h=1,2, \ldots, N/m$, the algorithm
has stored on the disk an array $SA_{ext}$ containing the sequence of
the first $(h-1)m$ suffixes of $T$ ordered lexicographically and
represented via their starting positions in $T$.\/}

During the $h$th stage, the algorithm {\em incrementally updates}
$SA_{ext}$ by properly inserting into it the text suffixes which start
in the substring $T[(h-1)m+1,hm]$. This preserves the invariant above,
thus ensuring that after all the $N/m$ stages, it is $SA_{ext}=SA_T$.
We are therefore left with showing how the generic $h$th stage works.

In the $h$th stage, the text substring $T[(h-1)m+1,hm]$ is loaded into
internal memory, and the suffix array $SA_{int}$ containing only the
suffixes starting in that text substring is built. Then, $SA_{int}$ is
merged with the current $SA_{ext}$ in two steps with the help of a
counter array $C[1,m+1]$:

\begin{enumerate}  

\item The text $T$ is scanned rightwards and the lexicographic
position $p_i$ of each text suffix $T[i,N]$, with $1 \leq i \leq
(h-1)m$, is determined in $SA_{int}$ via a binary search. The entry
$C[p_i]$ is then incremented by one unit in order to record the fact
that $T[i,N]$ lexicographically lies between the $SA_{int}[p_i-1]$-th
and the $SA_{int}[p_i]$-th suffix of $T$.

\item The information kept in the array $C$ is employed to quickly
merge $SA_{int}$ with $SA_{ext}$: entry $C[j]$ indicates how many
consecutive suffixes in $SA_{ext}$ follow the $SA_{int}[j-1]$-th text
suffix and precede the $SA_{int}[j]$-th text suffix. This implies that
a simple disk scan of $SA_{ext}$ is sufficient to perform such a
merging process. 

\end{enumerate}

At the end of these two steps, the invariant on $SA_{ext}$ has been
properly preserved so that $h$ can be incremented and the next stage
can start correctly. Some comments are in order at this point. It is
clear that the algorithm proceeds by mainly executing two disk scans:
one is performed to load the text piece $T[(h-1)m+1,hm]$ in internal
memory, the other disk scan is performed to merge $SA_{int}$ and
$SA_{ext}$ via the counter array $C$. However, the algorithm might
incur in many I/Os: either when $SA_{int}$ is built or when the
lexicographic position $p_i$ of each text suffix $T[i,N]$ within
$SA_{int}$ has to be determined. In both these two cases, we may need
to compare a pair of text suffixes which share a long prefix not
entirely available in internal memory (i.e., it extends beyond
$T[(h-1)m+1,hm]$). In the pathological case $T=a^N$, the comparison
between two text suffixes takes $O(N/M)$ bulk I/Os so that: $O(N\log_2
m)$ bulk I/Os are needed to build $SA_{int}$; the computation of $C$
takes $O(h N \log_2 m)$ bulk I/Os; whereas $O(h)$ bulk I/Os are needed
to merge $SA_{int}$ with $SA_{ext}$.  No random I/Os are executed, and
thus the global number of bulk I/Os is $O((N^3 \log_2 M) /M^2)$.  The
total space occupancy is $4N$ bytes for $SA_{ext}$ and $8m$ bytes for
both $C$ and $SA_{int}$; plus $m$ bytes to keep $T[(h-1)m+1,hm]$ in
internal memory (the value of $\ell$ is derived consequently). The
merging step can be easily implemented using some extra space (indeed
additional $4N$ bytes are sufficient), or by employing just the space
allocated for $SA_{int}$ and $SA_{ext}$ via a more tricky
implementation.

Since the worst-case number of total I/Os is cubic, a purely
theoretical analysis would classify this algorithm not much
interesting. But there are some considerations that are crucial to
shed new light on it, and look at this algorithm from a different
perspective. First of all, we must observe that, in practical
situations, it is very reasonable to assume that each suffix
comparison finds in internal memory all the (usually, constant number
of) characters needed to compare the two involved suffixes.
Consequently, the practical behavior is more reasonably described by
the formula: $O(N^2/M^2)$ bulk I/Os.  Additionally, in the analysis
above all I/Os are {\em sequential} and the actual number of random
seeks is $O(N/M)$ (i.e., at most a constant number per stage).
Consequently, the algorithm takes fully advantage of the large
bandwidth of current disks and of the high CPU-speed of the
processors~\cite{quantum,Ruemmler-Wilkes}.  Moreover, the reduced
working space facilitates the prefetching and caching policies of the
underlying operating system and finally, a careful look to the
algebraic calculations shows that the constants hidden in the big-Oh
notation are very small.  A recent result~\cite{CF99} has also shown
how to make it no longer questionable at theoretical eyes by proposing
a modification that achieves efficient performance in the worst case.

\medskip
{\bf From $SA_\Delta$ to $SBT_\Delta$. } The construction of
$SA_\Delta$ can be coupled with the computation of the array
$Lcp_\Delta$ containing the sequence of longest-common-prefix lengths
(lcp) between any pair of adjacent suffixes. Given these two arrays,
the String B-tree for the text collection $\Delta$ can be easily
derived proceeding in a bottom-up fashion.  We split $SA_\Delta$ into
groups of about $2b$ suffix pointers each (a similar splitting is
adopted on the array $Lcp_\Delta$) and use them to form the leaves of
the String B-tree.  That requires scanning $SA_\Delta$ and
$Lcp_\Delta$ once.  For each leaf $\pi$ we have its string set
$\S_\pi$ and its sequence of lcps, so that the construction of the
Patricia tree $PT_\pi$ takes linear time and no I/Os.

After the leaf level of the String B-tree has been constructed, we
proceed to the next higher level by determining new string and lcp
sequences.  For this, we scan rightward the leaf level and take the
leftmost string $L(\pi)$ and the rightmost string $R(\pi)$ from each
leaf $\pi$.  This gives the new string sequence whose length is a
factor $\Theta(1/B)$ smaller than the sequence of strings stored in
the leaf level. Each pair of adjacent strings is either a
$L(\pi)/R(\pi)$ pair or a $R(\pi)/L(\pi')$ pair (derived from
consecutive leaves $\pi$ and $\pi'$). In the former case, the lcp of
the two strings is obtained by taking the minimum of all the lcps
stored in $\pi$; in the latter case, the lcp is directly available in
the array $Lcp_\Delta$ since $R(\pi)$ and $L(\pi')$ are contiguous
there. After that the two new sequences of strings and lcps have been
constructed, we repeat the partitioning process above thus forming a
new level of internal nodes of the String B-tree. The process
continues for $O(\log_B N)$ iterations until the string sequence has
length smaller than $2b$; in that case the root of the String B-tree
is formed and the construction process stopped.  The implementation is
quite standard and not fully detailed here.  Preliminary
experiments~\cite{FG:96} have shown that the time taken to build a
String B-tree from its suffix array is negligible with respect to the
time taken for the construction of the suffix array itself. Hence we
refer the reader to~\cite{CF99} for the latter timings.

We conclude this section by observing that if we aim for optimal
I/O-bounds then we have to resort a suffix tree construction
method~\cite{FFM:00} explicitly designed to work in external
memory. The algorithm is too much sophisticated to be detailed, we
therefore refer the reader to the corresponding literature and, just,
point out here that the two arrays $SA_\Delta$ and $Lcp_\Delta$ can be
obtained from the suffix tree by means of an inorder traversal. It can
be shown that all these steps require sorting and sequential disk-scan
procedures, thus accounting for overall $O((N/B) \; \log_{M/B} (N/B))$
I/Os~\cite{FFM:00}.

\subsection{String vs suffix sorting} 
\label{string-suffix}

The construction of full-text indexes involves the sorting of the
suffixes of the indexed text collection. Since a suffix is a string of
arbitrary length, we would be driven to conclude that suffix sorting
and string sorting are ``similar'' problems. This is not true because,
intuitively, the suffixes participating to the sorting process share
so long substrings that some I/Os may be possibly saved when comparing
them, and indeed this saving can be achieved as shown theoretically
in~\cite{FFM:00}. Conversely~\cite{AFGV97} showed that sorting strings
on disk is not nearly as simple as it is in internal memory, and
introduced a bunch of sophisticated, deterministic string-sorting
algorithms which achieve I/O-optimality under some conditions on the
string-comparison model. In this section we present a simpler
randomized algorithm that comes close to the I/O-optimal complexity,
and surprisingly matches the $O(N/B)$ linear I/O-bound under some
reasonable conditions on the problem parameters.

Let $K$ be the number of strings to be sorted, they are arbitrarily
long, and let $N$ be their total length. For the sake of presentation,
we introduce the notation $n=N/B, k=K/B$ and $m=M/B$.  Since
algorithms do exist that match the $\Omega(K \log_2 K + N)$ lower
bound for string sorting in the comparison model, it seems reasonable
to expect that the complexity of sorting strings in external memory is
$\Theta(k \log_m k + n)$ I/Os. But any na\"{\i}ve algorithm does not
even come close to meet this I/O-bound. In fact, in internal memory a
trie data structure suffices to achieve the optimal complexity;
whereas in external-memory the use of the powerful String B-tree
achieves $O(K \log_B K + n)$ I/Os.  The problem here is that strings
have variable length and their brute-force comparisons over the
sorting process may induce a lot of I/Os. We aim at speeding up the
string comparisons, and we achieve this goal by {\em shrinking the
long strings via an hashing of some of their pieces}. Since hashing
does not preserve the lexicographic order, we will orchestrate the
selection of the string pieces to be hashed with a carefully designed
sorting process so that the correct sorted order may be eventually
computed. Details follow, see Figure~\ref{fig:random} for the
pseudocode of this algorithm.

\begin{figure}
\hrule\medskip  

{\small

\begin{description}

\item[Input:] A set $\S$ of $K$ strings, whose total length is $N$
(bits)

\item[Output:] A sorted permutation of $\S$

\end{description}

\begin{enumerate} 

\item Every string of $\S$ is partitioned into pieces of $L$ bits
each. $L$ is chosen to be much larger than $2 \log_2 K$.

\item Compute for each string piece a {\em name}, i.e. a bit string of
length $2 \log_2 K$, by means of a proper hash function. Each string of
$\S$ is then compressed by replacing $L$-pieces with their
corresponding names. The resulting set of compressed strings is
denoted with $\C$, and its elements are called {\em c-strings}.

\item Sort $\C$ via any known external-memory sorting algorithm
(e.g. Mergesort).

\item Compute the {\em longest common prefix} between any pair of
c-strings adjacent in (the sorted) $\C$ and mark the (at most two)
{\em mismatching names}. Let $lcp_x$ be the number of names shared by
the $x$th and the $(x+1)$th string of $\C$.

\item Scan the set $\C$ and collect the (two) marked names of each
c-string together with their corresponding $L$-pieces. Sort these
string pieces (they are at most $2K$) and assign a {\em rank} to each
of them--- equal pieces get the same rank. The rank is represented
with $2\log_2 K$ bits (like the names of the string pieces), possibly
padding the most significant digits with zeros.

\item Build a (logical) table $\T$ by mapping c-strings to columns and
names of L-pieces to table entries: $\T[a,b]$ contains the $a$th name
in the $b$th c-string of $\C$. Subsequently, transform $\T$'s entries
as follows: replace the marked names with their corresponding ranks,
and the other names with a bit-sequence of $2\log_2 K$ zeros.  If the
c-strings have not equal length, pad {\em logically} them with
zeros. This way names and ranks are formed by the same number of
bits, c-strings have the same length, and their (name or rank) pieces
are correctly aligned.

\item Perform a forward and backward pass through the columns of $\T$
as follows:

\begin{enumerate}

\item In the rightward pass, copy the first $lcp_{x-1}$ entries of the
$(x-1)$th column of $\T$ into the subsequent $x$th column, for
$x=2,...,K$. The mismatching names of the $x$th column are not
overridden.

\item In the leftward pass, copy the first $lcp_{x}$ entries of the
$(x+1)$th column of $\T$ into the $x$th column, for $x=K-1,....,1$.

\end{enumerate}

\item The columns of $\T$ are sorted via any known external-memory
sorting algorithm (e.g. Mergesort). From the bijection: {\em string}
$\leftrightarrow$ {\em c-string} $\leftrightarrow$ {\em column}; we
derive the sorted permutation of $\S$.

\end{enumerate} 
} 
\hrule

\caption{A randomized algorithm for sorting arbitrary long strings in
external memory.\label{fig:random}}
\end{figure}

We illustrate the behavior of the algorithm on a running example and
then sketch a proof of its correctness. Let $\S$ be a set of six
strings, each of length 10. In Figure~\ref{step1-4} these strings are
drawn vertically, divided into pieces of $L=2$ characters each. The
hash function used to assign names to the $L$-pieces is depicted in
Figure~\ref{names}.  We remark that $L \gg 2\log_2 K$ in order to
ensure, with high probability, that the names of the (at most $2K$)
mismatching $L$-pieces are different. Our setting $L=2$ is to simplify
the presentation.

\begin{figure}
\centerline{\begin{tabular}{|c|c|c|}
\hline
\mbox{ $L$-piece } & \mbox{ name }& \mbox{ rank }\\
\hline
\mbox{  aa  }& \mbox{  6   }& \mbox{  1  }\\
ab & 1 & 2\\
bb & 4 & 3\\
bc & 2 & -\\
ca & 5 & 4\\
cb & 3 & 5\\
cc & 7 & 6\\
\hline
\end{tabular}}
\caption{Names of all $L$-pieces and ranks of the marked
$L$-pieces. Notice that the L-piece $bc$ has no rank because it has
been not marked in Step~4.\label{names}}
\end{figure}

Figure~\ref{step1-4} illustrates the execution of Steps~1--4: from the
naming of the $L$-pieces to the sorting of the c-strings and finally
to the identification of the mismatching names.  We point out that
each c-string in $\C$ has actually associated a pointer to the
corresponding $\S$'s string, which is depicted in Figure~\ref{step1-4}
below every table; this pointer is exploited in the last Step~8 to
derive the sorted permutation of $\S$ from the sorted table
$\T$. Looking at Figure~\ref{step1-4}(iii), we interestingly note that
$\C$ is different from the sorted set $\S$ (in $\C$ the 4th string of
$\S$ precedes its 5th string~!), and this is due to the fact that the
names do not reflect of course the lexicographic order of their
original string pieces. The subsequent steps of the algorithm are then
designed to take care of this apparent disorder by driving the
c-strings to their correctly-ordered positions.

\begin{figure}
\centerline{
\begin{tabular}{|c|c|c|c|c|c|}
\hline
\mbox{ ab } & \mbox{ bb } & \mbox{ ab } & \mbox{ bb } & \mbox{ aa } & \mbox{ ab } \\
ab & bc & ab & bc & bb & cc \\
bc & ca & bc & cc & cc & aa \\
cb & aa & aa & cc & bb & bc \\
ab & bb & bb & aa & aa & ab \\
\hline 
\multicolumn{1}{c}{1} & \multicolumn{1}{c}{2} & \multicolumn{1}{c}{3} & \multicolumn{1}{c}{4} & \multicolumn{1}{c}{5} & \multicolumn{1}{c}{6} \\
\multicolumn{6}{c}{i. Step 1}
\end{tabular}
\hspace*{.3cm}
\begin{tabular}{|c|c|c|c|c|c|}
\hline
\mbox{  1  } & \mbox{  4   } & \mbox{  1  } & \mbox{  4  } & \mbox{  6  } & \mbox{  1  }\\
1 & 2 & 1 & 2 & 4 & 7 \\
2 & 5 & 2 & 7 & 7 & 6 \\
3 & 6 & 6 & 7 & 4 & 2 \\
1 & 4 & 4 & 6 & 6 & 1 \\
\hline 
\multicolumn{1}{c}{1} & \multicolumn{1}{c}{2} & \multicolumn{1}{c}{3} & \multicolumn{1}{c}{4} & \multicolumn{1}{c}{5} & \multicolumn{1}{c}{6} \\
\multicolumn{6}{c}{ii. Step 2}
\end{tabular}
\hspace*{.3cm}
\begin{tabular}{|c|c|c|c|c|c|}
\hline
\mbox{  1  }& \mbox{  1  }& \mbox{ {\underline 1} }& \mbox{ {\underline 4} }& \mbox{ {\underline 4} }& \mbox{ {\underline 6} }\\
1 & {\underline 1} & {\underline 7} & 2 & 2 & 4 \\
2 & 2 & 6 & {\underline 5} & {\underline 7} & 7 \\
{\underline 3} & {\underline 6} & 2 & 6 & 7 & 4 \\
1 & 4 & 1 & 4 & 6 & 6 \\
\hline 
\multicolumn{1}{c}{1} & \multicolumn{1}{c}{3} & \multicolumn{1}{c}{6} & \multicolumn{1}{c}{2} & \multicolumn{1}{c}{4} & \multicolumn{1}{c}{5} \\
\multicolumn{6}{c}{iii. Steps 3--5}
\end{tabular}
}
\caption{Strings are written from the top to the bottom of each table
column. (i) Strings are divided into pieces of 2 chars each. (ii) Each
$L$-piece is substituted with its name taken from the (hash) table of
Figure~\protect\ref{names}. (iii) Columns are sorted and mismatching
names between adjacent columns are underlined. \label{step1-4}}
\end{figure}

Step~6 builds the logical table $\T$ by substituting marked names with
their ranks (assigned in Step~5 and detailed in Figure~\ref{names}),
and the other names with zeros. Of course this transformation is {\em
lossy} because we have lost a lot of c-string characters (e.g. the
piece $bc$ which was not marked), nonetheless we will show below that
the canceled characters would have not been compared in sorting the
$\S$'s strings so that their eviction has not impact on the final
sorting step.  Figure~\ref{step6-8}(i-ii) shows how the forward and
backward scanning of table $\T$ fills some of its entries that got
zeros in Step~6. In particular Step~7(a) does not change table $\T$,
whereas Step~7(b) changes the first two columns. The resulting table
$\T$ is finally sorted to produce the correct sequence of string
pointers {\tt 5,3,1,6,2,4} (Figure~\ref{step6-8}(iii)).

\begin{figure}
\centerline{
\begin{tabular}{|c|c|c|c|c|c|}
\hline
\mbox{  0  }& \mbox{  0  }& \mbox{  2  }& \mbox{  3  }& \mbox{  3  }& \mbox{  1 } \\
0 & 2 & 6 & 0 & 0 & 0 \\
0 & 0 & 0 & 4 & 6 & 0 \\
5 & 1 & 0 & 0 & 0 & 0 \\
0 & 0 & 0 & 0 & 0 & 0 \\
\hline 
\multicolumn{1}{c}{1} & \multicolumn{1}{c}{3} & \multicolumn{1}{c}{6}
& \multicolumn{1}{c}{2} & \multicolumn{1}{c}{4} & \multicolumn{1}{c}{5} \\
\multicolumn{6}{c}{i. Step 6 and 7(a)}
\end{tabular}
\hspace*{.3cm}
\begin{tabular}{|c|c|c|c|c|c|}
\hline
\mbox{  2  }& \mbox{  2  }& \mbox{  2  }& \mbox{  3  }& \mbox{  3  }& \mbox{  1 } \\
2 & 2 & 6 & 0 & 0 & 0 \\
0 & 0 & 0 & 4 & 6 & 0 \\
5 & 1 & 0 & 0 & 0 & 0 \\
0 & 0 & 0 & 0 & 0 & 0 \\
\hline 
\multicolumn{1}{c}{1} & \multicolumn{1}{c}{3} & \multicolumn{1}{c}{6}
& \multicolumn{1}{c}{2} & \multicolumn{1}{c}{4} & \multicolumn{1}{c}{5} \\
\multicolumn{6}{c}{ii. Step 7(b)}
\end{tabular}
\hspace*{.3cm}
\begin{tabular}{|c|c|c|c|c|c|}
\hline
\mbox{  1  }& \mbox{  2  }& \mbox{  2  }& \mbox{  2  }& \mbox{  3  }& \mbox{  3 } \\
0 & 2 & 2 & 6 & 0 & 0 \\
0 & 0 & 0 & 0 & 4 & 6 \\
0 & 1 & 5 & 0 & 0 & 0 \\
0 & 0 & 0 & 0 & 0 & 0 \\
\hline 
\multicolumn{1}{c}{5} & \multicolumn{1}{c}{3} & \multicolumn{1}{c}{1}
& \multicolumn{1}{c}{6} & \multicolumn{1}{c}{2} & \multicolumn{1}{c}{4} \\
\multicolumn{6}{c}{iii. Step 8}
\end{tabular}
}
\caption{(i) The rightward pass through table $\T$. (ii) The leftward
pass through table $\T$. (iii) The sorted $\T$.\label{step6-8}}
\end{figure}

As far as the I/O-complexity is concerned, we let $sort(\eta,\mu)$
denote the I/O-cost of sorting $\eta$ strings of total length $\mu$
via multiway Mergesort, actually $sort(\eta,\mu) = O(\frac{\mu}{B} \;
\log_m \frac{\mu}{B})$.  Since the string set $\S$ is sequentially
stored on disk, Steps~1-2 take $O(n)$ I/Os. Step~3 sorts $K$ c-strings
of total length $N' = \Theta(\frac{N(2 \log_2 K)}{L}+K)$, where the
second additive term accounts for those strings which are shorter than
$L$, thus requiring $sort(K,N')$ I/Os.  Step~4 marks two names per
c-string, so Step~5 requires $sort(2K,2KL)$ I/Os. Table $\T$ consists
of $K$ columns of total length $N'$ bits. Hence, the forward and
backward scanning of Step~7 takes $O(N'/B)$ I/Os. Sorting the columns
of table $\T$ takes $sort(K,N')$ I/Os in Step~8. Summing up we have

\begin{theorem}
\label{teo:random}

The randomized algorithm detailed in Figure~\protect\ref{fig:random}
sorts $K$ strings of total length $N$ in $sort(K,N'+ 2KL) + n$
expected I/Os.

\end{theorem}

By setting $L=\Theta(\log_m{n} \; \log_2{K})$, the cost is $O(n + k
(\log_m{n})^2 \; \log_2{K})$ I/Os.  Moreover if it is $K \leq
N/(\log_m^2{n} \: \log_2{K})$, i.e. the average string length is
polylogarithmic in $n$, then the total sorting cost results the
optimal $O(n)$ I/Os.

It goes without saying that if one replaces the mismatching names with
their original $L$-pieces (instead of their ranks), it would still get
the correct lexicographic order but it would possibly end up in the
same I/O-cost of classical mergesort: in the worst case, Step~7
expands all entries of $\T$ thus resorting to a string set of size
$N$~!

\medskip
The argument underlying the proof of correctness of this algorithm is
non trivial. The key point is to prove that given any pair of strings
in $\S$, the corresponding columns of $\T$ (i.e. c-strings of $\C$)
contain enough information after Step~7 that the column comparison in
Step~8 reflects their correct lexicographic order. For simplicity we
assume to use a perfect hash function so that different $L$-pieces get
different names in Step~2.

Let $\alpha$ and $\beta$ be any two c-strings of $\C$ and assume that
they agree up to the $i$th name (included). After $\C$ is sorted
(Step~3), $\alpha$ and $\beta$ are possibly separated by some
c-strings which satisfy the following two properties: (1)~all these
c-strings agree at least up to their $i$th name, (2)~at least two
adjacent c-strings among them disagree at their $(i+1)$th
name. According to Step~6 and Property~(1), the columns in $\T$
corresponding to $\alpha$ and $\beta$ will initially get zeros in
their first $i$ entries; according to Step~6 and Property~(2) at least
two columns between $\alpha$'s and $\beta$'s ones will get a rank
value in their $(i+1)$th entry. The leftmost of these ranks equals the
rank of the $(i+1)$th name of $\alpha$; the rightmost of these ranks
equals the rank of the $(i+1)$th name of $\beta$. After Step~7, the
first $i$ entries of $\alpha$'s and $\beta$'s columns will be filled
with equal values; and their $(i+1)$th entry will contain two distinct
ranks which correctly reflect the two $L$-pieces occupying the
corresponding positions.  Hence the comparison executed in Step~8
between these two columns gives the correct lexicographic order
between the two original strings. Of course this argument holds for
any pair of c-strings in $\C$, and thus overall for all the columns of
$\T$. We can then conclude that the string permutation derived in
Step~8 is the correct one.

\subsection{Some open problems and future research directions}
\label{open-fulltext}

An important advantage of String B-trees is that they are a variant of
B-trees and consequently most of the technological advances and
know-how acquired on B-trees can be smoothly applied to them. For
example, split and merge strategies ensuring good page-fill ratio,
node buffering techniques to speed up search operations, B-tree
distribution over multi-disk systems, as well adaptive overflow
techniques to defer node splitting and B-tree re-organization, can be
applied on String B-trees without any significant
modification. Surprisingly enough, there are no publicly available
implementations of the String B-tree, whereas some softwares are based
on it~\cite{CFP99,divesh-nick,russi}. The novel ideas presented in this
paper foretell an engineered, publicly available implementation of
this data structure. In particular, it would be worth to design a
library for full-text indexing large text collections based on the
String B-tree data structure. This library should be designed to
follow the API of the Berkeley DB~\cite{berkeleyDB}, thus facilitating
its use in well-established applications.  The String B-tree could
also be adopted as the main search engine for genomic databases thus
competing with the numerous results based on suffix trees recently
appeared in the
literature~\cite{Gus97,chen97triebased,kahveci01,Heuman95,MS2000b}.
Another setting where an implementation of the String B-tree could
find a successful use is the indexing of the tagged structure of an
XML document. Recent results~\cite{fabric,muthu-twig-match,aboulnaga}
adopt a Patricia tree or a Suffix tree to solve and/or estimate the
selectivity of structural queries on XML documents. However they are
forced to either {\em summarize} the trie structure, in order to fit
it into the internal memory, or to propose disk-paging heuristics, in
order to achieve reasonable performance. Unfortunately these
proposals~\cite{fabric} forget the advancements in the string-matching
literature and thus inevitably incur into the well-known I/O
bottleneck deeply discussed in Section~\ref{sa-st}. Of course String
B-trees might be successfully used here to manage in an I/O-efficient
manner the arbitrary long XML paths in which an XML document can be
parsed, as well provide a better caching behavior for the in-memory
implementations.

The problem of multi-dimensional substring search, i.e. the search for
the simultaneous occurrence of $k$ substrings, deserves some
attention. The approach proposed in~\cite{pods01} provides some
insights into the nature of {\em two}-dimensional queries, but what
can we say about {\em multi}-dimensions~? Can we combine the String
B-tree with some known multi-dimensional data
structure~\cite{samet,GI99} in order to achieve guaranteed worst-case
bounds~?  Or, can we design a full-text index which allows proximity
queries between two substrings~\cite{manber-proximity,pods01}~? More
study is worth to be devoted to this important subject because of its
ubiquitous applications to databases, data mining and search engines.

When dealing with word-based indexes, we addressed the {\em document
listing problem}: given a word-based query $w$ find all the documents
in the indexed collection that contain $w$. Conversely when dealing
with full-text indexes, we addressed the {\em occurrence listing
problem}: given an arbitrary pattern string $P$ find all the document
positions where $P$ occurs. Although more natural from an
application-specific point of view, the document listing problem has
surprisingly received not much attention from the algorithmic
community in the area of full-text indexes, so that efficient
(optimal) solutions are yet missing for many of its variants.  Some
papers \cite{matias-muthu,muthu-soda02} have recently initiated the
study of challenging variations of the document listing problem and
solved them via simple and efficient algorithms. Improving these
approaches, as well extending these results to multiple-pattern
queries and to external-memory setting turns out to be a stimulating
direction of research.

\medskip
Exact searches are just one side of the coin, probably the tool with
the narrowest setting of application~! The design of search engines
for approximate or similarity string searches is becoming more urgent
because of the doubtless theoretical interest and the numerous
applications in the field of genomic databases, audio/video
collections and textual databases, in general. Significant biological
breakthroughs have already been achieved in genome research based on
the analysis of similar genetic sequences, and the algorithmic field
is overflooding of results in this
setting~\cite{navarro-survey}. However most of these similarity-based
or approximate-matching algorithms require the whole scan of the data
collection thus resulting much costly in the presence of a large
amount of string data and user queries. Indexes for approximate, or
similarity, searches turn out to be the holy grail of the Information
Retrieval field. Several proposals have appeared in the literature
and it would be impossible to comment the specialties of, or even
list, all of them. Just to have an idea, a search for ``(approximate
OR similarity) AND (index OR search)'' returned on Altavista more than
500,000 matches. To guide ourselves in this jungle of proposals we
state the following consideration: {\em ``it is not yet known an index
which efficiently routes the search to the correct positions where an
approximate/similar string occurrence lies''}. Most of the research
effort has been devoted to design {\em filters}: they transform the
approximate/similarity pattern search into another string or geometric
query problem for which efficient data structures are known. The
transformation is of course ``not perfect'' because it introduces some
{\em false positive} matches that must be then filtered out via a
(costly) scan-based algorithm. The more filtration is achieved by the
index, the smaller is the part on which the approximate/similar
scan-based search is applied, the faster is the overall algorithmic
solution. The key point therefore relies on the design of a good {\em
distance-preserving} transformation.

Some approaches transform the approximate search into a set of
$q$-gram exact searches, then solved via known full-text
indexes~\cite{esko92,quasar,navarro-qgram,joki-esko,pevzner,burkhardt-cpm02}.
Other approaches map a string onto a multi-dimensional integral point
via a {\em wavelet-based} transformation and then use
multi-dimensional geometric structures to solve the transformed
query~\cite{kahveci01}. Recently a more sophisticated
distance-preserving transformation has been introduced
in~\cite{muthu-stoc00,cormode-soda00} which maps a string into a
binary vector such that the {\em hamming distance} between two of
these vectors provides a provably good approximation of the {\em
(block) edit distance} between the two original strings. This way an
efficient approximate nearest-neighbor data structure (see
e.g.~\cite{indyk-stoc98,kushilevitz-stoc98}) can be used to search
over these multi-dimensional vectors and achieve guaranteed good
average-case performance. Notice that this solution applies on whole
strings; its practical performance has been tested over genomic data
in~\cite{cenk-muthu-cpm02}.

It goes without saying that in the plethora of results about complex
pattern searches a special place is occupied by the solutions based on
suffix trees~\cite{Gus97,st-approx,MS2000b,scozzesi}. The suffix-tree
structure is well suitable to perform regular expressions, approximate
or similarity-based searches but at an average-time cost which may be
exponential in the pattern length or polynomial in the text
length~\cite{navarro-survey}. Although some recent
papers~\cite{scozzesi,SV97,MS2000b} have investigated the
effectiveness of those results onto genomic databases, their
usefulness remains limited due to the I/O bottlenecks incurred by the
suffix tree both in the construction phase and for what concerns their
space occupancy (see Section~\ref{sa-st}). Perhaps the adaptation of
these complex searching algorithms to the String B-tree might turn
into appealing these approaches also from a practical perspective.

As a final remark, we mention that the techniques for designing
filtering indexes are not limited to genomic or textual databases, but
they may be used to extend the search functionalities of relational
and object-oriented databases, e.g. provide a support to {\em
approximate string joins}~\cite{muthu-approxjoins}. This shows a new
interesting direction of research for pattern-matching tools.

\medskip
In Section~\ref{word:caching} we addressed the problem of caching
inverted indexes for improving their query time under biased
operations. This issue is challenging over all the indexing schemes
and it becomes particularly difficult in the case of full-text indexes
because of their complicated structure. For example, in the case of a
suffix tree its unbalanced tree structure asks for an allocation of
its nodes to disk pages, usually called packing, that optimizes the
cache performance for some pattern of accesses to the tree nodes. This
problem has been investigated in~\cite{gil-itai} where an algorithm is
presented that finds an optimal packing with respect to both the total
number of different pages visited in the search and the number of page
faults incurred. It is also shown that finding an optimal packing
which minimizes also the space occupancy is, unfortunately,
NP-complete and an efficient approximation algorithm is
presented. These results deal with a static tree, so that it would be
interesting to explore the general situation in which the distribution
of the queries is not known in advance, changes over the time, and new
strings are inserted or deleted from the indexed set. A preliminary
insight on this challenging question has been achieved
in~\cite{ciriani-et-al}. There a novel self-adjusting full-text index
for external memory has been proposed, called SASL, based on a variant
of the Skip List data structure~\cite{Pugh90}. Usually a skip list is
turned into a {\em self-adjusting} data structure by {\em promoting}
the accessed items up its levels and {\em demoting} certain other
items down its levels~\cite{Cenk01,Mulmuley,Mehlhorn-Naher}. However
all of the known approaches fail to work effectively in an
external-memory setting because they lack locality of reference and
thus elicit a lot of random I/Os. A technical novelty of SASL is a
simple randomized demotion strategy that, together with a
doubly-exponential grouping of the skip list levels, guides the
demotions and guarantees {\em locality of reference} in all the
updating operations; this way, frequent items get to remain at the
highest levels of the skip list with high probability, and effective
I/O-bounds are achieved on expectation both for the search and update
operations.  SASL furthermore ensures balancedness without explicit
weight on the data structure; its update algorithms are simple and
guarantee a good use of disk space; in addition, SASL is with high
probability no worse than String B-trees on the search operations but
can be significantly better if the sequence of queries is highly
skewed or changes over the time (as most transactions do in
practice). Using SASL over a sequence of $m$ string searches
$S_{i_{1}},S_{i_{2}},\ldots,S_{i_{m}}$ takes
$O(\sum_{j=1}^{m}\left(\frac{|S_{i_{j}}|}{B}\right)+
\sum_{i=1}^{n}(n_{i}\log_{B}\frac{m}{n_{i}}))$ expected I/Os, where
$n_{i}$ is the number of times the string $S_{i}$ is queried.  The
first term is a lower bound for scanning the query strings; the second
term is the entropy of the query sequence and is a standard
information-theoretic lower bound. This is actually an extension of
the {\em Static Optimality Theorem} to external-memory string
access~\cite{Sleator-Tarjan}.

\medskip
In the last few years a number of models and techniques have been
developed in order to make it easier to reason about multi-level
hierarchies~\cite{Vitter}. Recently in~\cite{cache-oblivious1} it has
been introduced the elegant {\em cache-oblivious} model, that assumes
a two-level view of the computer memory but allows to prove results
for an unknown multilevel memory hierarchy. Cache oblivious algorithms
are designed to achieve good memory performance on all levels of the
memory hierarchy, even though {\em they avoid any memory-specific
parameterization}. Several basic problems--- e.g. matrix
multiplication, FFT,
sorting~\cite{cache-oblivious1,brodal-cache02b}--- have been solved
optimally, as well irregular and dynamic problems have been recently
addressed and solved via efficient cache-oblivious data
structures~\cite{farach-cache,brodal-cache02,bender-cache02}.  In this
research flow turns out challenging the design of a {\em cache
oblivious trie} because we feel that it would probably shed new light
on the indexing problem: it is not clear how to guarantee cache
obliviousness in a setting where items are arbitrarily long and the
size of the disk page is unknown.

%%%%%%%%%% COMPRESSION  %%%%%%%%%%%%%%%%%%%%%%%%%%%%%%%%%%%%%%%%%%

\section{Space-time tradeoff in index design}
\label{trade-off}

A {\em leitmotiv} of the previous sections has been the following:
{\em Inverted indexes occupy less space than full-text indexes but are
limited to efficiently support poorer search operations.} This is a
frequent statement in text indexing papers and talks, and it has
driven many authors to conclude that the increased query power of
full-text indexes {\em has to be paid} by additional storage
space. Although this observation is much frequent and apparently
established, it is challenging to ask ourselves if it is provable that
such a tradeoff {\em does exist} when designing an index.  In this
context compression appears as an attractive tool because it allows
not only to squeeze the space occupancy but also to improve the
computing speed. Indeed ``{\em space optimization is closely related
to time optimization in a disk memory\/}''~\cite{Knuth:1998:SS}
because it allows a better use of the fast and small memory levels
close to CPU (i.e. L1 or L2 caches), reduces the disk accesses,
virtually increases the disk bandwidth, and comes at a negligible cost
because of the significant speed of current CPUs. It is therefore not
surprising that IBM has recently installed on the eServers x330 a
novel memory chip (based on the Memory eXpansion
Technology~\cite{ibm-mxt}) that stores data in a compressed form thus
ensuring a performance similar to the one achieved by a server with
double real memory but, of course, at a much lower cost. All these
considerations have driven developers to state that it is {\em more
economical to store data in compressed form than uncompressed}, so
that a renewed interest in compression techniques raised within the
algorithmic and IR communities.

We have already discussed in Section~\ref{word-based} the use of
compression in word-based index design, now we address the impact of
compression onto full-text index design. 

Compression may of course operate at the text level or at the index
level, or both. The simplest approach consists of compressing the text
via a lexicographic-preserving code~\cite{hu-tucker} and then build a
suffix array upon it~\cite{MNZ97}. The improvement in space occupancy
is however negligible since the index is much larger than the text. A
most promising and sophisticated direction was initiated
in~\cite{Munro97,Munro98} with the aim of compressing the full-text
index itself. These authors showed how to build a suffix-tree based
index on a text $T[1,n]$ within $n \log_2 n + O(n)$ bits of storage
and support the search for a pattern $P[1,p]$ in $O(p+occ)$ worst-case
time. This result stimulated an active research on {\em succinct
encodings} of full-text indexes that ended up with a
breakthrough~\cite{GV:00} in which it was shown that a {\em succinct}
suffix array can be built within $\Theta(n)$ bits and can support
pattern searches in $O(\frac{p}{\log_2 n} + occ \log^{\epsilon} n)$
time, where $\epsilon$ is an arbitrarily small positive constant. This
result has shown that the apparently ``random'' permutation of the
text suffixes can be succinctly coded in optimal space in the worst
case~\cite{demaine-ortiz-soda01}. In~\cite{sada00,sada-soda02}
extensions and variations of this result--- e.g. an arbitrary large
alphabet--- have been considered.

The above index, however, uses space linear in the size of the indexed
collection and therefore it results not yet competitive against the
word-based indexes, whose space occupancy is usually $o(n)$ (see
Section~\ref{word-based}). Real text collections are compressible and
thus a full-text index should desiderably exploit the repetitiveness
present into them to squeeze its space occupancy via a much succinct
coding of the suffix pointers.  

The first step toward the design of a {\em truly compressed full-text
index} ensuring effective search performance in the worst case has
been recently pursued in~\cite{FM:00}.  The novelty of this approach
resides in the careful combination of the Burrows-Wheeler compression
algorithm~\cite{bw} with the suffix array data structure thus
designing a sort of {\em compressed suffix array}.  It is actually a
{\em self-indexing tool} because it encapsulates a compressed version
of the original text inside the compressed suffix array. Overall we
can say that the index is {\em opportunistic} in that, although no
assumption on a particular text distribution is made, it takes
advantage of the compressibility of the indexed text by decreasing the
space occupancy at no significant slowdown in the query
performance. More precisely, the index in~\cite{FM:00} occupies $O(n
\; H_k(T)) + o(n)$ bits of storage, where $H_k(T)$ is the $k$-th order
empirical entropy of the indexed text~$T$, and supports the search for
an arbitrary pattern $P[1,p]$ as a substring of $T$ in $O(p + occ
\log^{\epsilon} n)$ time.

In what follows we sketch the basic ideas underlying the design of
this compressed index, hereafter called \fmind~\cite{FM:00}, and we
briefly discuss some experimental results~\cite{FM:01,is01} on various
text collections.  These experiments show that the \fmind\ is compact
(its space occupancy is close to the one achieved by the best known
compressors), it is fast in counting the number of pattern
occurrences, and the cost of their retrieval is reasonable when they
are few (i.e. in case of a selective query). As a further contribution
we briefly mention an interesting adaptation of the \fmind\ to
word-based indexing, called \wfmind. This result highlights further on
the interplay between compression and index design, as well the recent
plot between word-based and full-text indexes: everything of these
worlds must be deeply understood in order to perform valuable research
in this topic.

\subsection{The Burrows-Wheeler transform} 
\label{bwt:back}

Let $T[1,n]$ denote a text over a finite alphabet
$\Sigma$. In~\cite{bw} Burrows and Wheeler introduced a new
compression algorithm based on a reversible transformation, now called
the {\em Burrows-Wheeler Transform} (BWT from now on). The BWT
permutes the input text $T$ into a new string that is easier to
compress.  The BWT consists of three basic steps (see
Figure~\ref{bwtfig}): (1)~append to the end of $T$ a special character
$\#$ smaller than any other text character; (2)~form a {\em logical}
matrix $\M$ whose rows are the cyclic shifts of the string $T\#$
sorted in lexicographic order; (3)~construct the transformed text $L$
by taking the last column of $\M$. Notice that every column of $\M$,
hence also the transformed text $L$, is a permutation of $T\#$.  In
particular the first column of $\M$, call it $F$, is obtained by
lexicographically sorting the characters of $T\#$ (or, equally, the
characters of $L$). The transformed string $L$ usually contains long
runs of identical symbols and therefore can be efficiently compressed
using move-to-front coding, in combination with statistical coders
(see for example~\cite{bw,fen96}).

\begin{figure}[t]
\newcommand{\xx}{\#}
\begin{center}
{\tt
\begin{tabular}{l}
                \\ \hline
mississippi\xx  \\
ississippi\xx m \\
ssissippi\xx mi \\
sissippi\xx mis \\
issippi\xx miss \\
ssippi\xx missi \\
sippi\xx missis \\
ippi\xx mississ \\
ppi\xx mississi \\
pi\xx mississip \\
i\xx mississipp \\
\xx mississippi \\
\hline
\end{tabular}
\makebox[1.5cm]{$\Longrightarrow$}
\begin{tabular}{l@{\hspace{0.15cm}}l@{\hspace{0.15cm}}l}
{\rm F}  &              &{\rm L} \\ \hline
\xx &  mississipp &i \\
i & \xx mississip &p \\
i & ppi\xx missis &s \\
i & ssippi\xx mis &s \\
i & ssissippi\xx  &m \\
m & ississippi &\xx  \\
p & i\xx mississi &p \\
p & pi\xx mississ &i \\
s & ippi\xx missi &s \\
s & issippi\xx mi &s \\
s & sippi\xx miss &i \\
s & sissippi\xx m &i \\
\hline
\end{tabular}
}
\end{center}
\caption{Example of Burrows-Wheeler transform for the string $T = {\tt
mississippi}$. The matrix on the right has the rows sorted in lexicographic
order. The output of the BWT is column $L$; in this example the string {\tt
ipssm\xx pissii}.}\label{bwtfig}
\end{figure}

\subsection{An opportunistic index}
\label{sub:opp}

There is a bijective correspondence between the rows of $\M$ and the
suffixes of $T$ (see Figure~\ref{bwtfig}); and thus there is a strong
relation between the string $L$ and the suffix array built on
$T$~\cite{MM93}.  This is a crucial observation for the design of the
\fmind. We recall below the basic ideas underlying the search
operation in the \fmind, referring for the other technical details to
the seminal paper~\cite{FM:00}.

In order to simplify the presentation, we distinguish between two
search tools: the counting of the number of pattern occurrences in $T$
and the retrieval of their positions.  The counting is implemented by
exploiting two nice structural properties of the matrix $\M$:
$(i)$~all suffixes of $T$ prefixed by a pattern $P[1,p]$ occupy a
contiguous set of rows of $\M$ (see also Section~\ref{sa-st});
$(ii)$~this set of rows has starting position $first$ and ending
position $last$, where $first$ is the {\em lexicographic position} of
the string $P$ among the ordered rows of $\M$. The value $(last -
first + 1)$ accounts for the total number of pattern occurrences. For
example, in Figure~\ref{bwtfig} for the pattern $P={\tt si}$ we have
$first = 9$ and $last = 10$ for a total of two occurrences.

The retrieval of the rows $first$ and $last$ is implemented by the
procedure \countx\ which takes $O(p)$ time in the worst case, working
in $p$ constant-time phases numbered from $p$ to 1 (see the pseudocode
in Fig.~\ref{search_alg}). Each phase preserves the following
invariant: {\em At the $i$-th phase, the parameter ``first'' points to
the first row of $\M$ prefixed by $P[i,p]$ and the parameter ``last''
points to the last row of $\M$ prefixed by $P[i,p]$}. After the final
phase, $first$ and $last$ will delimit the rows of $\M$ containing all
the text suffixes prefixed by $P$.

\begin{figure}
\hrule {\algskip \alg{Algorithm $\countx(P[1,p])$}
\begin{enumerate}

\item $i = p$, $\ch = P[p]$, $first = C[\ch]+1$, $last = C[\ch+1]$;

\item {\bf while} $((first \leq last)
\mbox{\bf \ \  and  } (i \geq 2))$ {\bf do}

\item \hspace*{1cm} $\ch = P[i-1]$;

\item \hspace*{1cm} $first = C[\ch] + \num(\ch,first-1)+1$;\label{upd:sp}

\item \hspace*{1cm} $last = C[\ch] + \num(\ch,last)$;\label{upd:ep}

\item \hspace*{1cm} $i = i-1$;

\item {\bf if} $(last < first)$
{\bf then return} ``no rows prefixed by $P[1,p]$'' {\bf else return}
$(first,last)$.

\end{enumerate}
\vskip0pt}\hrule
\caption{Algorithm \countx\ finds the set of rows prefixed by pattern
$P[1,p]$. Procedure $\num(\ch,k)$ counts the number of occurrences of
the character $\ch$ in the string prefix $L[1,k]$. In~\cite{FM:00} it
is shown how to implement $\num(\ch,k)$ in constant
time.\label{search_alg}}
\end{figure}

The location of a pattern occurrence is found by means of algorithm
\locate. Given an index $i$, $\locate(i)$ returns the starting
position in $T$ of the suffix corresponding to the $i$th row in
$\M$. For example in Figure~\ref{bwtfig} we have $pos(3)=8$ since the
third row ${\tt ippi\# mississ}$ corresponds to the suffix $T[8,11] =
{\tt ippi}$.  The basic idea for implementing $\locate(i)$ is the
following. We {\it logically mark} a suitable subset of the rows of
$\M$, and for each marked row $j$ we store the starting position
$pos(j)$ of its corresponding text suffix. As a result, if
\locate($i$) finds the $i$th row marked then it immediately returns
its position $pos(i)$; otherwise, \locate\ uses the so called {\em
LF-computation} to move to the row corresponding to the suffix
$T[pos(i)-1,n]$. Actually, the index of this row can be computed as
$LF[i]=C[L[i]] + \num(L[i],i)$, where $C[c]$ is the number of
occurrences in $T$ of the characters smaller than $c$. The
LF-computation is iterated $v$ times until we reach a marked row $i_v$
for which $pos(i_v)$ is available; we can then set $pos(i) = pos(i_v)
+ v$. Notice that the LF-computation is considering text suffixes of
increasing length, until the corresponding marked row is encountered.

\medskip
Given the appealing asymptotical performance and structural properties
of the \fmind, the authors have investigated in~\cite{FM:01,is01} its
practical behavior by performing an extensive set of experiments on
various text collections: 1992~CIA world fact book (shortly {\it
world}) of about 2Mb, King James Bible (shortly {\it bible}) of about
4Mb, DNA sequence (shortly {\it e.coli}) of about 4Mb, SGML-tagged
texts of AP-news (shortly, {\em ap90}) of about 65Mb, the java
documentation (shortly, {\it jdk13}) of about 70Mb, and the Canterbury
Corpus (shortly, {\em cantrbry}) of about 3Mb. On these files they
actually experimented two different implementations of the \fmind:

\begin{itemize}

\item A {\em tiny} index designed to achieve high compression but
supporting only the counting of the pattern occurrences.

\item A {\em fat} index designed to support both the counting and the
retrieval of the pattern occurrences.

\end{itemize}

Both the tiny and the fat indexes consist of a compressed version of
the input text plus some additional information used for pattern
searching. In Table~\ref{zip_table} we report a comparison among these
compressed full-text indexes, {\gzip} (the standard Unix compressor)
and \bzip\ (the best known compressor based on the
BWT~\cite{bzip2_home}). These figures have been derived
from~\cite{is01,FM:01}.

\begin{table}
\setlength{\tabcolsep}{0.16cm}
\begin{center}\small
\begin{tabular}{|l|l|r|r|r|r|r|r|}\hline
\multicolumn{2}{|r|}{File}
       &{\it bible}&{\it e.coli}&{\it world}&{\it cantbry}&{\it jdk13}&
        {\it ap90}\\ \hline
{\sf tiny index} &
Compr. ratio & 21.09 & 26.92 & 19.62 & 24.02 & 5.87 & 22.14 \\
 &
Construction time & 2.24 & 2.19 & 2.26 & 2.21 & 3.43 & 3.04\\
 &
Decompression time & 0.45 & 0.49 & 0.44 & 0.38 & 0.42 & 0.57 \\
 &
Ave. \countxx\ time& 4.3  &  12.3 & 4.7 &  8.1& 3.2  & 5.6 \\\hline
{\sf fat index} &
Compr. ratio & 32.28 & 33.61& 33.23& 46.10& 17.02& 35.49\\
 &
Construction time & 2.28 & 2.17 & 2.33 & 2.39 & 3.50 & 3.10\\
 &
Decompression time & 0.46 & 0.51 & 0.46 & 0.41 & 0.43 & 0.59\\
 &
Ave. \countxx\ time& 1.0  &  2.3 & 1.5 &  2.7& 1.3  & 1.6 \\
 &
Ave. \locate\ time& 7.5  & 7.6  & 9.4 &  7.1& 21.7 & 5.3   \\\hline
\bzip &
Compression ratio& 20.90 & 26.97 & 19.79 & 20.24 & 7.03 & 27.36\\
 &
Compression time& 1.16 & 1.28 & 1.17 & 0.89 & 1.52 & 1.16 \\
 &
Decompression time & 0.39 & 0.48 & 0.39 & 0.31 & 0.28 & 0.43 \\\hline
\gzip\ &
Compr. ratio & 29.07 & 28.00 & 29.17 & 26.10 & 10.79 & 37.35 \\
 &
Compression time & 1.74 & 10.48 & 0.87 & 5.04 & 0.39 & 0.97 \\
 &
Decompression time & 0.07 &  0.07 & 0.06 & 0.06 & 0.04 & 0.07 \\ \hline
\end{tabular}
\end{center}
\caption{Compression ratio (percentage) and compression/decompression
speed (microseconds per input byte) of tiny and fat indexes compared
with those of \gzip\ (with option {\tt -9} for maximum compression)
and \bzip.  For these compressed indexes we also reports the average
time (in milliseconds) for the \countxx\ and \locate\ operations.  The
experiments were run on a machine equipped with Gnu/Linux Debian~2.2,
600Mhz Pentium~III and 1~Gb RAM.\label{zip_table}}
\end{table}

The experiments show that the tiny index takes significantly less
space than the corresponding \gzip-compressed file, and for all files
except {\it bible} and {\it cantrbry} it takes less space than
\bzip. This may appear surprising since \bzip\ is also based on the
BWT~\cite{bzip2_home}. The explanation is simply that the \fmind\
computes the BWT for the entire file whereas \bzip\ splits the input
in 900Kb blocks. This compression improvement is payed in terms of
speed; the construction of the tiny index takes more time than \bzip.
The experiments also show that the fat index takes slightly more space
than the corresponding \gzip-compressed file. For what concerns the
query time we have that both the tiny and the fat index compute the
number of occurrences of a pattern in a few milliseconds,
independently of the size of the searched file. Using the fat index
one can also compute the position of each occurrence in a few
milliseconds per occurrence.

These experiments show that the \fmind\ is compact (its space
occupancy is close to the one achieved by the best known compressors),
it is fast in counting the number of pattern occurrences, and the cost
of their retrieval is reasonable when they are few (i.e. in case of a
selective query). In addition, the \fmind\ allows to trade space
occupancy for search time by choosing the amount of auxiliary
information stored into it.  As a result the \fmind\ combines
compression and full-text indexing: like \gzip\ and \bzip\ it
encapsulates a compressed version of the original file; like suffix
trees and arrays it allows to search for arbitrary patterns by looking
only at a small portion of the compressed file.

\subsection{A word-based opportunistic index}
\label{wfmind}

As far as user queries are formulated on arbitrary substrings, the
\fmind\ is an effective and compact search tool. In the information
retrieval setting, thought, user queries are commonly {\em word-based}
since they are formulated on entire words or on their parts, like
prefixes or suffixes. In these cases, the \fmind\ suffers from the
same drawbacks of classical full-text indexes: at any word-based query
formulated on a pattern $P$, it needs a {\em post-processing} phase
which aims at filtering out the occurrences of $P$ which are not word
occurrences because they lie entirely into a text word.  This mainly
consists of checking whether an occurrence of $P$, found via the
\countx\ operation, is {\em preceded and followed by a non-word
character}.  In the presence of frequent query-patterns such a
filtering process may be very time consuming, thus slowing down the
overall query performance. This effect is more dramatic when the goal
is to {\em count} the occurrences of a word, or when we need to just
{\em check} whether a word does occur or not into an indexed text.

Starting for these considerations the \fmind\ has been enriched with
some additional information concerning with the {\em linguistic
structure} of the indexed text.  The new data structure, called
\wfmind, is actually obtained by building the \fmind\ onto a {\em
``digested''} version of the input text. This digested text, shortly
$DT$, is a special compressed version of the original text $T$ that
allows to map {\em word-based queries on $T$ onto substring queries on
$DT$}.

More precisely, the digested text $DT$ is obtained by compressing the
text $T$ with the {\em byte-aligned} and {\em tagged} Huffword
algorithm described in Section~\ref{word-based}
(see~\cite{NMN00}). This way $DT$ is a {\em byte sequence} which
possesses a crucial property: {\em Given a word $w$ and its
corresponding tagged codeword $cw$, we have that $w$ occurs in $T$ iff
$cw$ occurs in $DT$}. The tagged codewords are in some sense {\em
self-synchronizing at the byte level} because of their most
significant bit set to 1. In fact it is not possible that a
byte-aligned codeword overlaps two or more other codewords, since it
should have at least one internal byte with its most significant bit
set to 1. Similarly, it is not possible that a codeword is
byte-aligned and starts inside another codeword, because the latter
should again have at least one internal byte with its most significant
bit set to 1.  Such a bijection allows us to convert every {\em
word-based query} formulated on a pattern $w$ and the text $T$, into a
{\em byte-aligned substring query} formulated on the tagged codeword
$cw$, relative to $w$, and the digested text $DT$.

Of course more complicated word queries on $T$, like prefix-word or
suffix-word queries, can be translated into {\em multiple substring}
queries on $DT$ as follows. Searching for the occurrences of a pattern
$P$ as a prefix of a word in $T$ consists of three steps: (1)~search
in the Huffword dictionary $\D$ for all the words prefixed by $P$, say
$w_1, w_2, \ldots, w_k$; (2)~compute the tagged codewords $cw_1, cw_2,
\ldots, cw_k$ for these words, and then (3)~search for the occurrences
of the $cw_i$ into the digested text $DT$. Other word-based queries
can be similarly implemented.

It is natural to use an \fmind\ built over $DT$ to support the
codeword searches over the digested text. Here the \fmind\ takes as
characters of the indexed text $DT$ its constituting {\em bytes}. This
approach has a twofold advantage: it reduces the space occupied by the
(digested) byte sequence $DT$ and supports over $DT$ effective
searches for byte-aligned substrings (i.e. codewords).

The \wfmind\ therefore consists of two parts: a full-text index
\fmind($\D$) built over the Huffword dictionary $\D$, and a full-text
index \fmind($DT$) built over the digested text $DT$. The former index
is used to search for the queried word (or for its variants) into the
dictionary $\D$; from the retrieved words we derive the corresponding
(set of) codewords which are then searched in $DT$ via
\fmind($DT$). Hence a single word-based query on $T$, can be
translated by \wfmind\ into a set of {\em exact substring} queries to
be performed by \fmind($DT$).

The advantage of the \wfmind\ over the standard \fmind\ should be
apparent. Queries are word-oriented so that the time consuming
post-processing phase has been avoided; counting or existential
queries are directly executed on the (small) dictionary $\D$ without
even accessing the compressed file; the overall space occupancy is
usually smaller than the one required by the \fmind\ because $\D$ is
small and $DT$ has a lot of structure that can be exploited by the
Burrows-Wheeler compressor present in \wfmind. This approach needs
further experimental investigation and engineering, although some
preliminary experiments have shown that \wfmind\ is very promising.

\subsection{Some open problems and future research directions}
\label{open-tradeoff}

In this section we have discussed the interplay between data
compression and indexing. The \fmind\ is a promising data structure
which combines effective space compression and efficient full-text
queries.  Recently, the authors of~\cite{FM:00} have shown that
another compressed index does exist that, based on the BWT and the
Lempel-Ziv parsing~\cite{lz78}, answers arbitrary pattern queries in
$O(p+occ)$ time and occupies $O(n\hk{k}(T)\log^\epsilon n) + o(n)$
bits of storage. Independently, \cite{navarro-lz} has presented a
simplified compressed index that does not achieve these good
asymptotic bounds but it could be suitable for practical
implementation. The main open problem left in this line of research is
the design of a data structure which achieves the best of the previous
bounds: $O(p+occ)$ query time and $O(n\hk{k}(T)) + o(n)$ bits of
storage occupancy.  However, in our opinion, the most challenging
question is if, and how, locality of reference can be exploited in
these data structures to achieve efficient I/O-bounds. We aim at
obtaining $O(occ/B)$ I/Os for the location of the pattern occurrences,
where $B$ is the disk-page size. In fact, the additive term $O(p)$
I/Os is negligible in practice because any user-query is commonly
composed of few characters. Conversely $occ$ might be large and thus
force the \locate\ procedure to execute many random I/Os in the case
of a large indexed text collection. An I/O-conscious compressed index
might compete successfully against the String B-tree data structure
(see Section~\ref{sb-tree}).

The Burrows-Wheeler transform plays a central role in the design of
the \fmind. Its computation relies on the construction of the suffix
array of the compressed string; this is the actual algorithmic
bottleneck for a fast implementation of this compression
algorithm. Although a plethora of papers have been devoted to
engineering the suffix sorting
step~\cite{bw,szip,fen96,bzip2_home,S98:COSTR,nel96,bs}, there is
still room for improvement~\cite{FM:02SA} and investigation. Any
advancement in this direction would immediately impact on the
compression time performance of \bzip.  As far as the compression
ratio of \bzip\ is concerned, we point out that the recent
improvements presented in the literature are either limited to special
data collections or they are not fully
validated~\cite{chapin,chapin-tate,S98:VAR,S99:MOD,fen96,kurtz99,kurtz00}.
Hence the open-source software \bzip\ yet remains {\em the}
choice~\cite{bzip2_home}. Further study, simplification or variation
on the Burrows-Wheeler transform are needed to improve its compression
ratio and/or possibly impact on the design of new compressed
indexes. The approach followed in \wfmind\ is an example of this line
of research.

\medskip
Although we have explained in the previous sections how to perform
simple exact searches, full-text indexes can do much more. In
Section~\ref{sa-st} we have mentioned that suffix trees can support
complex searches like approximate or similarity-based matches, as well
regular expression searches. It is also well-known that suffix arrays
can simulate any algorithm designed on suffix trees at an $O(\log n)$
extra-time penalty. This slowdown is payed for by the small space
occupied by the suffix array. It is clear at this point that it should
be easy to adapt these algorithms to work on the \fmind\ or on the
\wfmind. The resulting search procedures might benefit more from the
compactness of these indexes, and therefore possibly turn into
in-memory some (e.g. genomic) computations which now require the use
of disk, with consequent poor performance. This line of research has
been pioneered in the experimental setting by~\cite{sadakane-approx}
which showed that compressed suffix arrays can be used as {\em
filtering data structure} to speed up similarity-based searches on
large genomic databases. From the theoretical point of view,
\cite{crochemore-soda02} recently proposed another interesting use of
compression for speeding up similarity-based computations in the worst
case. There the dynamic programming matrix has been divided into
variable sized blocks, as induced by the Lempel-Ziv parsing of both
strings~\cite{lz78}, and the inherent periodic nature of the strings
has been exploited to achieve $O(n^2/\log n)$ time and space
complexity. It would be interesting to combine these ideas with the
ones developed for the \fmind\ in order to reduce the space
requirements of these algorithms without impairing their sub-quadratic
time complexity (which is conjectured in~\cite{crochemore-soda02} to
be close to optimal).

\medskip
The \fmind\ can also be used as a building block of sophisticated
Information Retrieval tools. In Section~\ref{word-based} we have
discussed the block-addressing scheme as a promising approach to index
moderate sized textual collections, and presented some approaches to
combine compression and block-addressing for achieving better
performance~\cite{glimpse,NMN00}. In these approaches opportunistic
string-matching algorithms have been used to perform searches on the
compressed blocks thus achieving an improvement of about 30-50\% in
the final performance.  The \fmind\ and \wfmind\ naturally fit in this
framework because they can be used to index each text block
individually~\cite{FM:00}; this way, at query time, the compressed
index built over the candidate blocks could be employed to fasten the
detection of the pattern occurrences. It must be noted here that this
approach fully exploits one of the positive properties of the
block-addressing scheme: {\em The vocabulary allows to turn complex
searches on the indexed text into multiple exact-pattern searches on
the candidate text blocks}. These are properly the types of searches
efficiently supported by \fmind\ and \wfmind. A theoretical
investigation using a model generally accepted in Information
Retrieval~\cite{CIKM97_jou} has showed in~\cite{FM:00} that this
approach achieves {\em both sublinear space overhead and sublinear
query time independent of the block size}. Conversely, inverted
indices achieve only the second goal~\cite{wmb99}, and the classical
Glimpse tool achieves both goals but under some restrictive conditions
on the block size~\cite{CIKM97_jou}. Algorithmic engineering and
further experiments on this novel IR system are yet missing and worth
to be pursued to validate these good theoretical results.

%%%%%%%%%% CONCLUSIONS  %%%%%%%%%%%%%%%%%%%%%%%%%%%%%%%%%%%%%%%%%%

\section{Conclusions}
\label{concl}

In this survey we have focused our attention on algorithmic and data
structural issues arising in two aspects of information retrieval
systems design: (1)~representing textual collections in a form which
is suitable to efficient searching and mining; (2)~design algorithms
to build these representations in reasonable time and to perform
effective searches and processing operations over them. Of course this
is not the whole story about this huge field as the {\em Information
Retrieval} is. We then conclude this paper by citing other important
aspects that would deserve further consideration: (a)~file structures
and database maintenance; (b)~ranking techniques and clustering
methods for scoring and improving query results; (c)~computational
linguistics; (d)~user interfaces and models; (e)~distributed retrieval
issues as well security and access control management. Every one of
these aspects has been the subject of thousands of papers and
surveys~! We content ourselves to cite here just some good starting
points from which a user can browse for further technical deepenings
and bibliographic links~\cite{wmb99,baeza99,compling,sigir}.

\subsubsection*{Acknowledgments}
This survey is the outcome of hours of highlighting and, sometime hard
and fatiguing, discussions with many fellow researchers and
friends. It encapsulates some results which have already seen the
light in various papers of mine; some other scientific results,
detailed in the previous pages, are however yet unpublished and
probably they'll remain in this state! So I'd like to point out the
persons who participated to the discovery of those ideas. The
engineered version of String B-trees (Section~\ref{engineering}) has
been devised in collaboration with Roberto Grossi; the randomized
algorithm for string sorting in external memory
(Section~\ref{string-suffix}) is a joint result with Mikkel Thorup;
finally, the \wfmind\ (Section~\ref{wfmind}) is a recent advancement
achieved together with Giovanni Manzini. I finally thanks Valentina
Ciriani and Giovanni Manzini for carefully reading and commenting the
preliminary versions of this survey.

\end{document}